\RequirePackage{fix-cm}
\documentclass[smallextended]{svjour3}       
\smartqed  

\usepackage{comment}

\usepackage[graphicx]{realboxes}
\usepackage{url}
\usepackage{tcolorbox}
\usepackage{multirow}
\usepackage{amssymb,amsfonts}
\usepackage{pifont}
\usepackage{xcolor,colortbl} 
\usepackage{listings}
\usepackage{hyperref}
\usepackage[figuresright]{rotating}
\usepackage{rotating}
\usepackage{pdflscape}
\hypersetup{colorlinks,breaklinks,
	linkcolor=darkblue,urlcolor=darkblue,
	anchorcolor=darkblue,citecolor=darkblue}
\usepackage{natbib}
\setcitestyle{authoryear,open={(},close={)}}
\usepackage{marvosym}

\usepackage{algorithm}
\usepackage{amsmath}
\usepackage[noend]{algpseudocode}
\usepackage[tight,footnotesize]{subfigure}

\usepackage{xcolor, soul}
\colorlet{lightgray}{gray!30}
\sethlcolor{lightgray} 

\definecolor{darkgray}{gray}{0.78}
\definecolor{lightgray}{gray}{0.85}
\definecolor{verylightgray}{gray}{0.95}

\definecolor{lightgray}{gray}{0.85}
\definecolor{verylightgray}{gray}{0.95}
\definecolor{darkblue}{rgb}{0.1,0.6,1.0}
\definecolor{red}{rgb}{1.0, 0.01, 0.24}
\definecolor{gray}{rgb}{0.8, 0.8, 0.8}

\makeatletter
\def\BState{\State\hskip-\ALG@thistlm}
\makeatother

\usepackage{xspace}
\usepackage{amssymb}
\usepackage[colorinlistoftodos]{todonotes}
\usepackage{xcolor}

\newcommand*{\ie}{i.e.,\@\xspace}
\newcommand*{\eg}{e.g.,\@\xspace}

\newcommand*{\RM}{README\@\xspace}

\newcommand*{\GH}{GitHub\@\xspace}
\newcommand*{\ME}{\texttt{Metagente}\@\xspace}
\newcommand*{\GS}{\texttt{GitSum}\@\xspace}

\newcommand*{\CA}{\texttt{Prompt Creator Agent}\@\xspace}
\newcommand*{\EA}{\texttt{Extractor Agent}\@\xspace}
\newcommand*{\SA}{\texttt{Summarizer Agent}\@\xspace}
\newcommand*{\TA}{\texttt{Teacher Agent}\@\xspace}

\newcommand*{\Fo}{\texttt{GPT-4o}\@\xspace}

\makeatletter
\newcommand*{\etc}{%
	\@ifnextchar{.}%
	{etc}%
	{etc.\@\xspace}%
}
\makeatother

\definecolor{darkgray}{gray}{0.78}
\definecolor{lightgray}{gray}{0.85}
\definecolor{verylightgray}{gray}{0.95}
\definecolor{codegreen}{rgb}{0,0.6,0}
\definecolor{codegray}{rgb}{0.5,0.5,0.5}
\definecolor{codepurple}{rgb}{0.58,0,0.82}
\definecolor{backcolour}{rgb}{0.95,0.95,0.92}


\newcommand{\mybox}[4]{
	\begin{figure}[h]
		\centering
		\begin{tikzpicture}
			\node[anchor=text,text width=\columnwidth-0.5cm, draw, rounded corners, line width=1pt, fill=#3, inner sep=2mm] (big) {\\#4};
			\node[draw, rounded corners, line width=.3pt, fill=#2, anchor=west, xshift=3mm] (small) at (big.north west) {#1};
		\end{tikzpicture}
	\end{figure}
}

\lstdefinestyle{java}{
	backgroundcolor=\color{backcolour},   commentstyle=\color{codegreen},
	keywordstyle=\color{magenta},
	numberstyle=\tiny\color{codegray},
	stringstyle=\color{codepurple},
	basicstyle=\ttfamily\scriptsize,
	breakatwhitespace=false,         
	breaklines=true,                 
	captionpos=b,                    
	keepspaces=true,                 
	numbers=left,                    
	numbersep=5pt,                  
	showspaces=false,                
	showstringspaces=false,
	showtabs=false,                  
	tabsize=2
}

\newcommand{\rqfirst}{\textbf{RQ$_1$}: \emph{How does \ME perform compared to single LLM-based agents?}} 

\newcommand{\rqsecond}{\textbf{RQ$_2$}: \emph{Does the dynamic iteration strategy contribute to 
		training efficiency?}} 

\newcommand{\rqthird}{\textbf{RQ$_3$}: \emph{Which LLM contributes to a better performance of \ME?}}

\begin{document}

\title{\ME: An LLMs-based Multi-Agent System for the Summarization of \GH \RM Files}

\title{\ME: Empowering the Synergy Among LLMs for \GH \RM Summarization}

\title{\ME: Augmenting the Synergy Among LLMs for \GH \RM Summarization}

\title{\ME: Augmenting the Synergy Among Large Language Models for \RM Summarization}

\title{Turning \RM files into About Descriptions: An Approach with LLMs-based Multi-Agent Systems}

\title{Turning \RM files into About Descriptions: An LLMs-based Multi-Agent Approach}

\title{Turning Long \RM Files into Short About Descriptions: An LLMs-based Multi-Agent Approach}

\title{Automated Summarization of Software Documents: An LLM-based Multi-Agent Approach}

\titlerunning{An LLM-based MAS for the Summarization of Software Documents}        

\author{Duc S. H. Nguyen \and Minh T. Nguyen \and  Phuong T. Nguyen \and Juri Di Rocco \and Davide Di Ruscio}



\institute{
	Duc S. H. Nguyen \at
	Hanoi University of Science and Technology, Vietnam \\ 
	\email{duc.nsh231061m@sis.hust.edu.vn}
	\and
	Minh T. Nguyen \at
	Hanoi University of Science and Technology, Vietnam \\ 
	\email{minh.nt225450@sis.hust.edu.vn}
	\and
	Phuong T. Nguyen \at
	University of L'Aquila, Italy \\ 
	\email{phuong.nguyen@univaq.it}
	\and	
	Juri Di Rocco \at
	University of L'Aquila, Italy \\ 
	\email{juri.dirocco@univaq.it}
	\and
	\Letter~Davide Di Ruscio \at
	University of L'Aquila, Italy \\ 
	\email{davide.diruscio@univaq.it}
}


\maketitle

\begin{abstract}
	Large Language Models (LLMs) and LLM-based Multi-Agent Systems (MAS) are revolutionizing software engineering (SE) by advancing automation, decision-making, and knowledge processing. Their recent application to SE tasks has already shown promising results. In this paper, we focus on summarization as a key application area. We present \ME, an LLM-based MAS designed to generate concise and accurate summaries of software documentation. 
	\ME employs a Teacher–Student architecture where multiple LLM agents collaborate to enhance relevance and precision of produced summaries. 	An empirical evaluation on real-world datasets demonstrates \ME's effectiveness in streamlining workflows, outperforming the considered baselines. The evaluation provides evidence that \ME improves summarization for requirements analysis and technical documentation.
	Our findings underscore the transformative potential of these technologies in SE, while identifying challenges and future research directions for their seamless integration.

	\keywords{Summarization \and LLMs \and Multi-Agent Systems \and GitHub \and Google Play}
    
\end{abstract}

	\section{Introduction}
	\label{sec:Introduction}

In Software Engineering, there has been an increasing use of documents to provide detailed descriptions for software artifacts. Among others, \RM files in \GH are written in the Markdown format, containing information about a repository, \eg instructions, help, or 
updates. A well-written \RM facilitates reading comprehension, helping visitors to grasp the scope of a repository. 
This is also the case of mobile app stores like Google Play Store,\footnote{\url{https://play.google.com/store/apps}} in which apps are normally equipped with a long HTML document, allowing visitors to understand the apps' functionalities. 
An empirical study~\citep{LIU2022106924} showed that \RM is often the very first item that visitors take a look at 
when it comes to becoming acquainted with a \GH repository. Similarly, apps' descriptions are of high importance, as they can be used as sources to mine domain knowledge~\citep{LIU2017126}.

By several \GH repositories and Google Play apps, instruction files are usually lengthy, and reading and understanding them can
require time and effort, thus discouraging visitors from continuing with the repositories/apps.  
Therefore, apart from \RM, \GH also allows its users to add a short description called ``About'' to each repository, offering a succinct summary of the main functionalities. In a similar manner, besides a full \RM file, apps in Google Play Store feature a brief description, resembling the ``About'' summary in \GH repositories. 
Unfortunately, in these platforms, while being useful, this field is usually overlooked and left unfilled, posing difficulties for those who want to explore the repositories, on the fly. Recently, \GS~\citep{10.1145/3593434.3593448} has been proposed as the first approach to summarize long \RM files to yield brief but concise  ``About'' descriptions. Being built on top of BART and T5, \GS successfully produces relevant summaries starting from long \RM files. However, while obtaining an encouraging recommendation performance, the tool still suffers from a low accuracy for input files that are of mixed fields, including code and text. This triggers the need for a more effective way to summarize \RM files.

Large Language Models (LLMs) have been 
applied in 
Software Engineering~\citep{DBLP:journals/software/Ozkaya23b,NGUYEN2024112059} to solve a wide range of tasks. To name but a few, researchers have employed LLMs 
in code summarization~\citep{sun2024source,haldar2024analyzing}, debugging~\citep{lee2024github,tian2024debugbench}, testing~\citep{arawjo2024chainforge,li2025enhancing}, or 
code generation~\citep{gu2023llm,fakhoury2024llm,huang2024bias}. 
However, LLMs are not without limitations: Existing work~\citep{He_Treude_Lo_2024} has revealed that while they are gaining popularity, their performance on tasks requiring domain-specific expertise or complex multi-step reasoning remains limited. In this context, prompt-tuning is a practical approach, in which prompts are iteratively refined to improve the performance of the pre-trained language model without modifying its internal design. Noteworthy, the prompt-tuning process can be prone to subjective bias and scalability issues, making it difficult to generalize across diverse tasks \citep{white2023promptpatterncatalogenhance}. Moreover, it is not easily accessible for several developers as it might require advanced technical expertise. 

To cope with the limitations of single LLMs,
multi-agent systems have been proposed to enable specialized LLMs to collaborate within a shared framework~\citep{He_Treude_Lo_2024,DBLP:journals/fcsc/WangMFZYZCTCLZWW24}. These systems magnify the unique strengths of various LLMs, where agents specialize in tasks such as code generation, debugging, or domain-specific problem-solving \citep{He_Treude_Lo_2024,DBLP:journals/corr/abs-2407-01489}. Challenges such as effective coordination, efficient communication, and the overhead of integrating multiple agents persist, even though dedicated frameworks like \texttt{LangChain}\footnote{\url{https://www.langchain.com/}} or \texttt{LLamaIndex}\footnote{\url{https://www.llamaindex.ai/}} are now increasingly adopted in Software Engineering to mitigate such issues. Nevertheless, MASs represent a compelling alternative to relying solely on the capabilities of individual models~\citep{10.1145/3691620.3695291}.

In our previous work \citep{10.1145/3696630.3728511}, we developed 
\ME, a multi-agent framework composed of four LLM-based agents, which 
interact in a collaborative manner using a teacher-student loop, enabling prompt optimization with minimal supervision. 
An evaluation using datasets collected from \GH demonstrated that \ME obtains a good recommendation performance, showing that the proposed approach is suitable for generating summaries for software documents. 
In this paper, we enhance \ME across several dimensions. \emph{First}, to improve efficiency, we propose a dynamic iteration strategy that adaptively halts processing of low-potential samples, reducing computational cost without compromising quality. \emph{Second}, we extend the framework to incorporate multiple LLMs as the recommendation engine, enhancing its flexibility. \emph{Third}, we broaden the evaluation to include summarization of structured HTML-based application descriptions into concise texts. This task is particularly challenging due to the heterogeneous content of HTML files, which can complicate summarization. Our evaluation on curated datasets demonstrates that \ME achieves strong semantic alignment and competitive ROUGE scores while requiring fewer training iterations, outperforming various baselines.

In summary, the main contributions of this work include the novel dynamic iteration strategy, the integration of diverse LLMs, and the expanded evaluation scope for complex summarization tasks, elaborated as follows.
\begin{itemize}
    \item \textbf{Solution.} We introduced an extended version of \ME, a collaborative LLM-based multi-agent system for summarizing app descriptions. Aiming for efficiency, a dynamic stopping mechanism has been deployed, so as to optimize the fine-tuning process. 
    \item \textbf{Evaluation.} Using datasets collected for mobile apps' documents, we conducted an empirical evaluation to study the performance of \ME. Moreover, we also compared dynamic vs. non-dynamic strategies, evaluating generalization across test subsets.
    \item \textbf{Reproducibility.} The replication package including code, data, and prompts has been released to foster reproducibility.\footnote{\url{https://github.com/MDEGroup/Metagente}}
\end{itemize}

\noindent
\textbf{Structure.} Section~\ref{sec:Background} provides some background related to the importance of \RM and ``About'' descriptions in 
Google Play. Section~\ref{sec:Proposed_Approach} introduces \ME, our proposed approach to the summarization of \RM files using LLM-based MAS. The evaluation is elaborated in Section~\ref{sec:Evaluation}, and the results are reported and analyzed in Section~\ref{sec:Results}. In Section~\ref{sec:Discussion}, there are discussion on the impacts, and the threats to the validity of the findings. We review the related work in Section~\ref{sec:RelatedWork}. Finally, Section~\ref{sec:Conclusions} sketches future work, and concludes the paper.

	\section{Motivation and Background}	
	\label{sec:Background}

Section~\ref{sec:MotivatingExample} presents a motivating example to demonstrate the necessity of ``About'' descriptions in the Google Play ecosystem. Afterwards, the ROUGE scores are presented in Section~\ref{sec:Metrics} as a base for the presentation of \ME in Section~\ref{sec:Proposed_Approach}.

\subsection{Motivating Examples}
\label{sec:MotivatingExample}

\begin{figure*}[t!]
	\centering
	\begin{tabular}{c c}	
		\subfigure[Front page]{\label{fig:Telegram-Front}
			\includegraphics[width=0.35\linewidth]{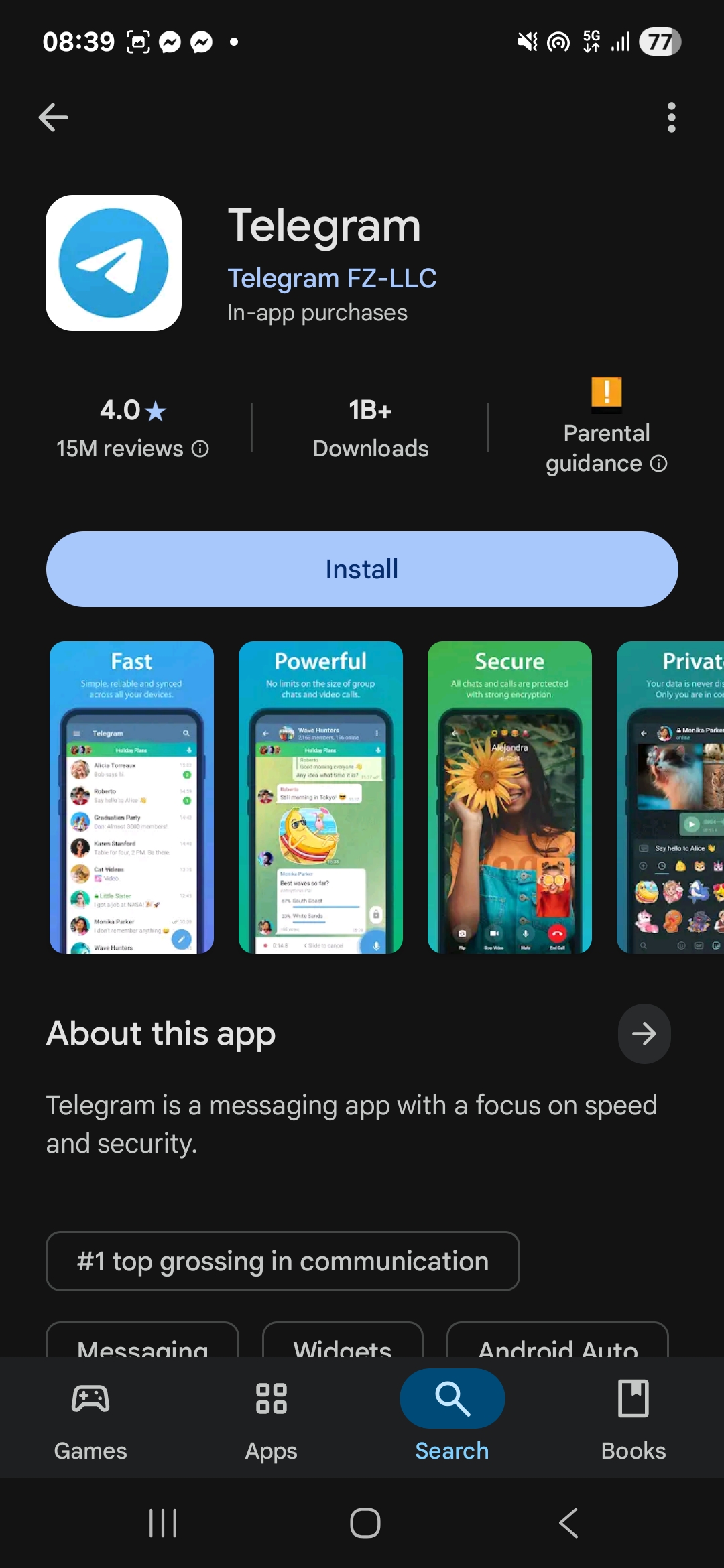}}	&
		\subfigure[README]{\label{fig:Telegram-Back}
			\includegraphics[width=0.35\linewidth]{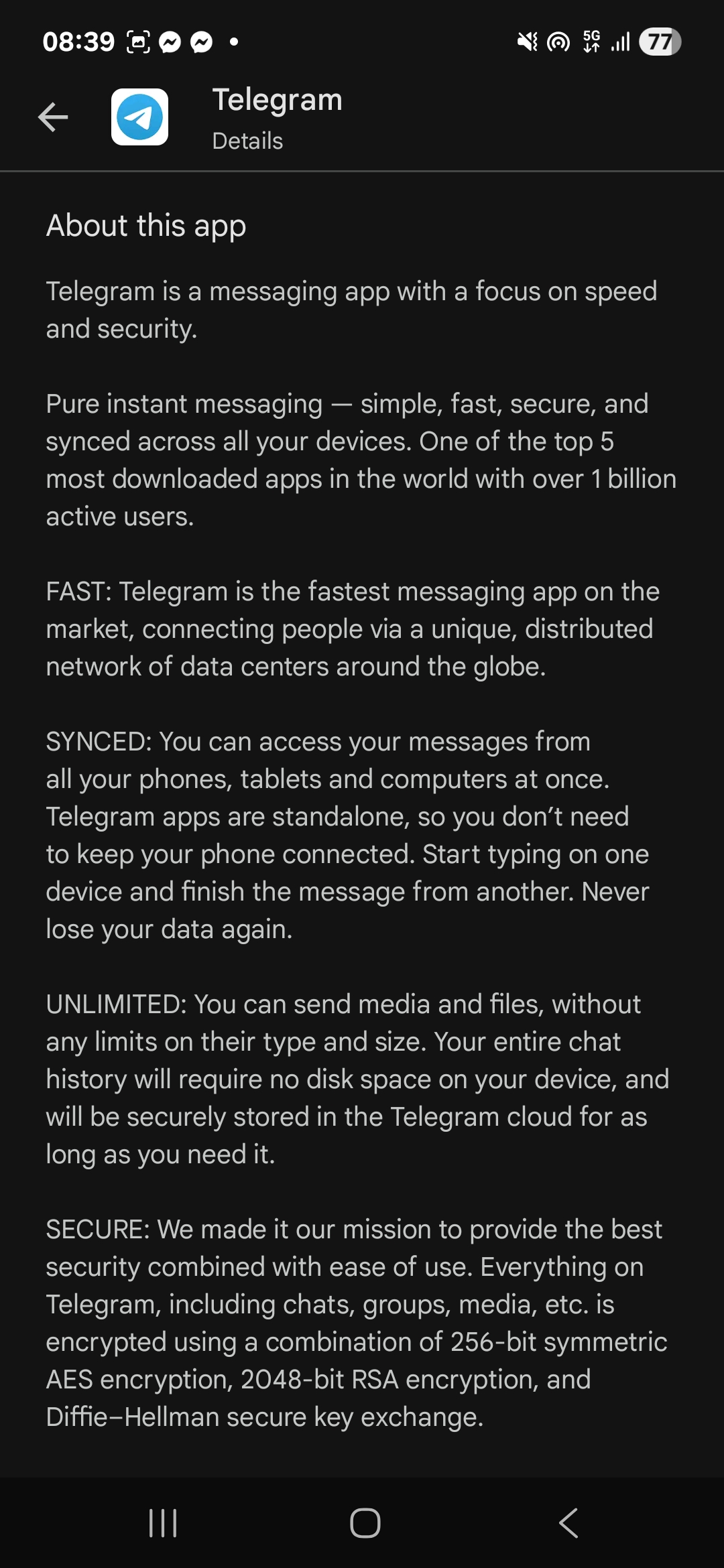}} 
	\end{tabular} 
	\caption{\texttt{Telegram} features a brief description for the app in its front page.} 	
	\label{fig:Telegram}
\end{figure*}

Figure~\ref{fig:Telegram-Front} shows 
the \texttt{Telegram} app, seen from the frontpage on Google Play Store. 
Apart from essential information including the number of reviews (15M with 4.0 as rating), number of downloads ($>$1B), 
there is also a short description, which is actually the ``About'' field, describing the main functionalities, \ie ``\emph{Telegram is a messaging app with focus on speed and security}.'' Such a description is brief but informative, and it helps 
users--especially first-time visitors--gain a quick orientation to decide whether to continue with the app, without reading the corresponding long \RM, which resides behind the frontpage shown in Figure~\ref{fig:Telegram-Back}. %

\begin{figure*}[t!]
	\centering
	\begin{tabular}{c c}	
		\subfigure[Front page]{\label{fig:Courtside1891-Front}
			\includegraphics[width=0.35\linewidth]{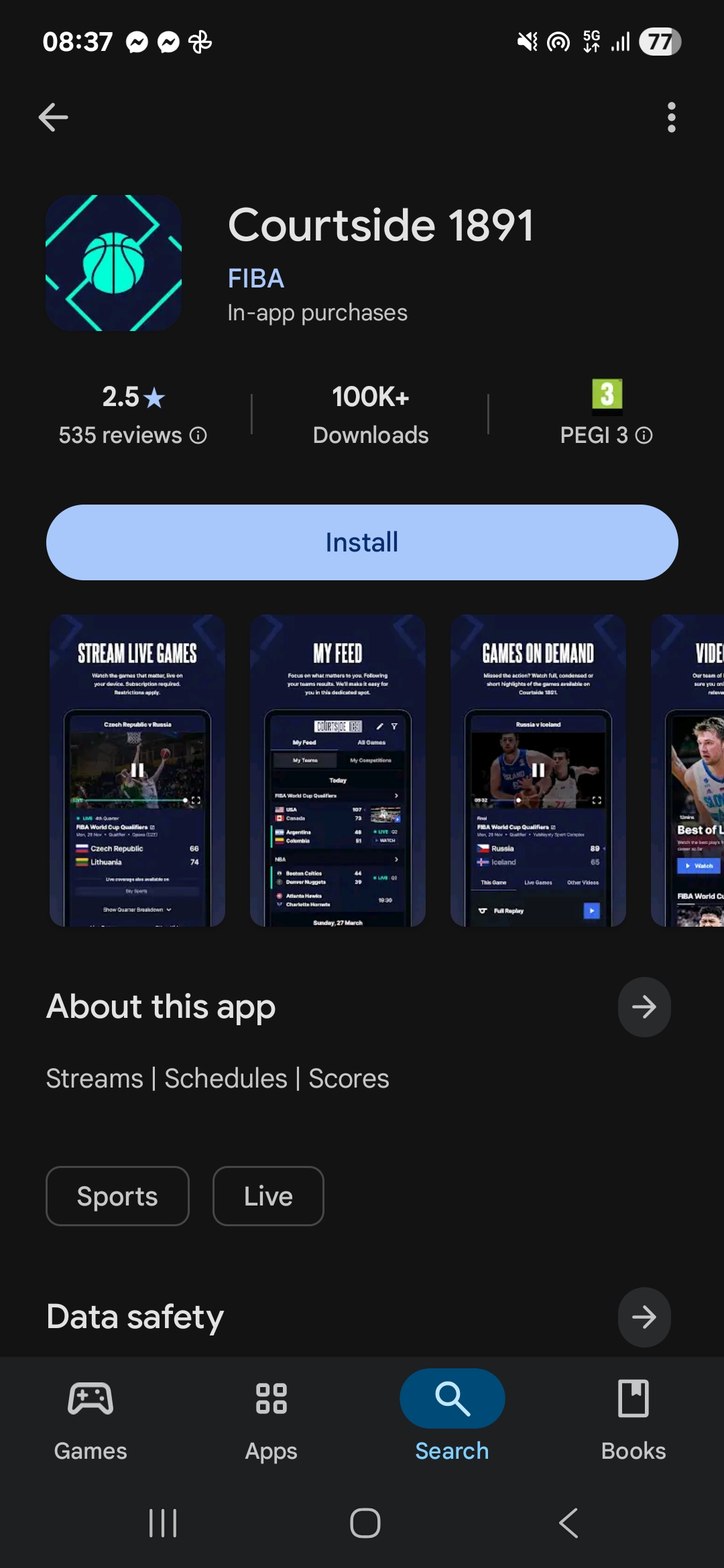}}	&
		\subfigure[README]{\label{fig:Courtside1891-Back}
			\includegraphics[width=0.35\linewidth]{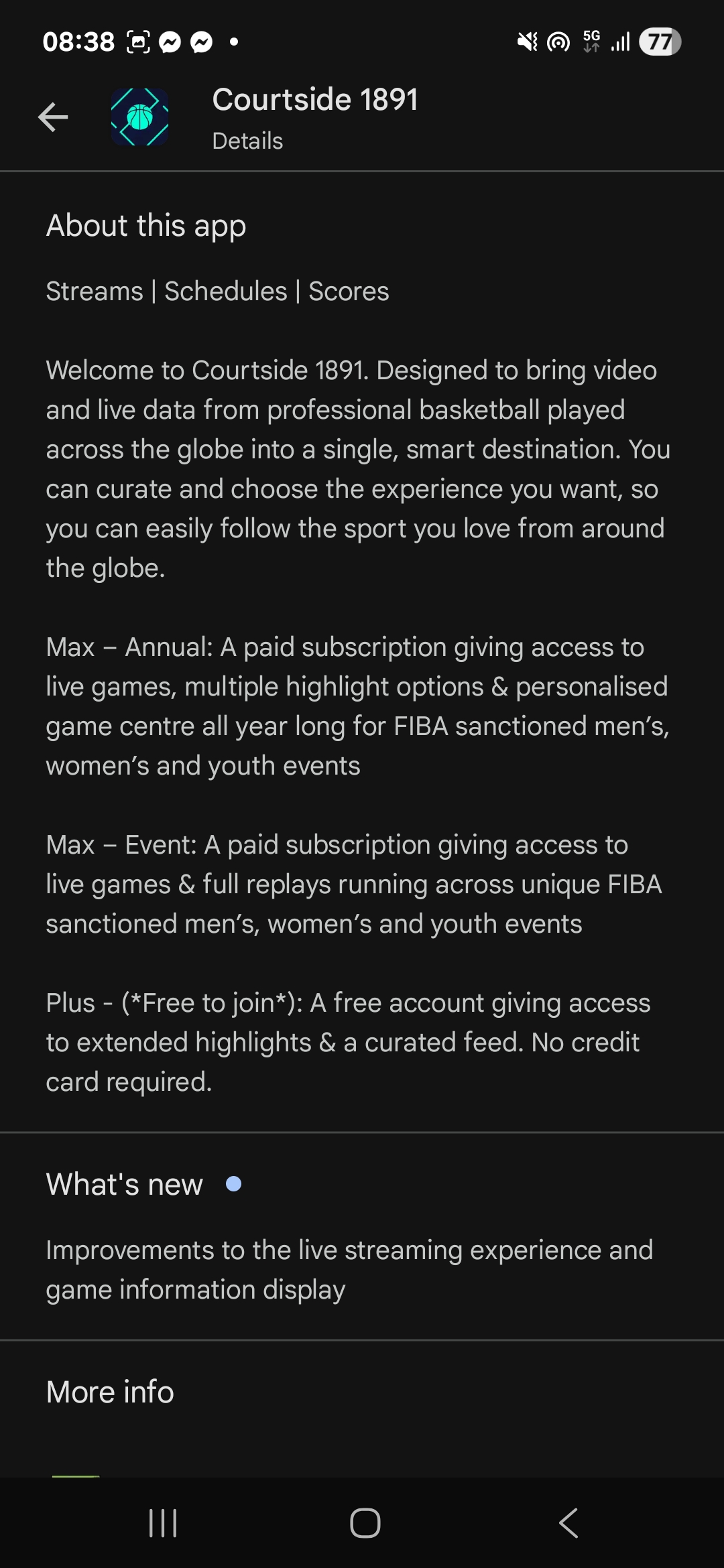}} 
	\end{tabular} 
	\caption{Despite a lengthy \RM file, \texttt{Courtside 1891} does not offer any informative descriptions.} 	
	\label{fig:Courtside1891}
\end{figure*}

Figure~\ref{fig:Courtside1891} depicts another example of 
\texttt{Courtside 1891}\footnote{\url{https://play.google.com/store/apps/details?id=com.pl.fiba&hl=en}} that is also a well-maintained app. However, while having a long and detailed \RM (see Figure~\ref{fig:Courtside1891-Back}), the app does not feature any short 
description (see Figure~\ref{fig:Courtside1891-Front}). Due to such a lack, visitors need to spend time reading the \RM in order to understand the app's functionalities. Essentially--in some extent--
this may discourage visitors from continuing with the app.

The examples motivate us to come up with an approach to summarize long \RM documents to yield a short description. In our previous work~\citep{10.1145/3593434.3593448}, we conceived \GS as a first approach to translate long \GH \RM files to obtain ``About'' descriptions. Using BART and T5 as the recommendation engine, \GS can generate  
relevant summaries for \GH repositories. 
However, while earning a high accuracy in different testing samples, \GS still struggles with input files composed of miscellaneous fields, including code and text.
In this paper, we aim to overcome the current limitations, leveraging the synergies of different task-specific LLMs that are properly orchestrated to enhance the recommendation quality. 
In particular, we employ an LLM to perform the extraction of the input data, filtering out noisy fields and retaining the most meaningful ones. More importantly, by means of a Teacher-Student architecture applying collaborative LLMs, we iteratively refine a set of prompts to produce the best fit prompt, which then can be used to generate the final summary. 

The next subsection describes the metrics used by \ME to steer the adjustment phase of LLMs-based agents.

\subsection{Metrics for Evaluating Summarization Tasks}\label{sec:Metrics}



Recall-Oriented Understudy for Gisting Evaluation (ROUGE) \citep{rouge2004} is a set of metrics applied in evaluating automatic summarization and machine translation 
\citep{Chen2020,itiger2022}. Given a pair of a produced summary 
and a ground-truth text, ROUGE metrics judge 
their similarity based on the overlap between them. 
ROUGE metrics are in the range of 0 and 1, in which a higher value corresponds to a higher similarity between the automatically produced summary and the reference one.

The metrics are computed as follows: 
\vspace{-.1cm}
\begin{equation*}
	\begin{aligned}
		Precision_{rouge-n} =&\frac{  \sum_{R,G \in S} \sum_{gram_n \in R} Count_G(gram_n)  }{   \sum_{R,G \in S} \sum_{gram_n \in G} Count_G(gram_n)   } \\
		Recall_{rouge-n}=&\frac{  \sum_{R,G \in S} \sum_{gram_n \in R} Count_G(gram_n)  }{   \sum_{R,G \in S} \sum_{gram_n \in R} Count_R(gram_n)   } \\
		F1_{rouge-n}=&\frac{2* Precision_{rouge-n} *  Recall_{rouge-n}}{Precision_{rouge-n} + Recall_{rouge-n}} 
	\end{aligned}
\end{equation*}
where $R$, $G$, and $S$ are the reference summary, generated summary, and the test set, respectively. $gram_n$ is an n-gram phase, where $n$ is the length of a word sequence; $Count_R(gram_n)$ and $Count_G(gram_n)$ are the occurrence number of $gram_n$ in $R$ and $G$. 



ROUGE-1 and ROUGE-2 are calculated with $N=1,2$, 
\ie uni-gram and bi-grams. $\sum_{gram_n \in R} Count_G(gram_n)$ represents the number of N-grams existing in both the ground-truth summary and generated summary. The aforementioned equations are interpreted as follows. Given two sets of N-grams produced from a generated summary and its original summary, respectively, Precision evaluates the proportion of N-grams in the first set that is found in the second set, while Recall measures the proportion of N-grams in the second set that exists in the first set. 
ROUGE F1-score is based on Longest Common Subsequence (LCS) defined as follows: 

\begin{equation*}
	\begin{aligned}
		Precision_{rouge-l}=&\frac{LCS(R,G)}{length(G)};        Recall_{rouge-l}=&\frac{LCS(R,G)}{length(R)} \\
		F1_{rouge-l}=&\frac{2* Precision_{rouge-l} *  Recall_{rouge-l}}{Precision_{rouge-l} + Recall_{rouge-l}} 
	\end{aligned}
\end{equation*}
where $LCS(R,G)$ represents the length of the longest common subsequence of 
reference summary and its generated summary ($R$ and $G$), respectively. ROUGE-L indicates the natural similarity between the two given sequences in sentence-level structure. It is also considered to be useful in evaluating summarization performance. 

In our proposed approach, ROUGE scores are used as means to guide the learning process, allowing the recommendation engine to refine the summaries, making them more relevant. The detailed architecture  of \ME together with its constituent components is described in the next section.



%
	
	\section{Proposed Approach}
	\label{sec:Proposed_Approach}


This section introduces the \ME approach to the summarization of software documents using LLMs-based agents. 
To deal with noise in the input data, an LLM has been deployed as an extractor component, 
filtering out noisy and retaining the most meaningful fields. More importantly, by means of a Teacher-Student architecture applying collaborative LLMs, we iteratively refine a set of prompts to produce the best fit prompt, which then can be used to generate the final summary. As shown 
in Fig.~\ref{fig:pipeline_combined}, there are two main phases, \ie Optimization and Evaluation, explained as follows.


\subsection{Optimization} 

\begin{figure*}[t!]
	\centering    
	\begin{tabular}{c}		
		\subfigure[Optimization Pipeline]{\label{fig:opt_pipeline}\includegraphics[width=0.88\textwidth]{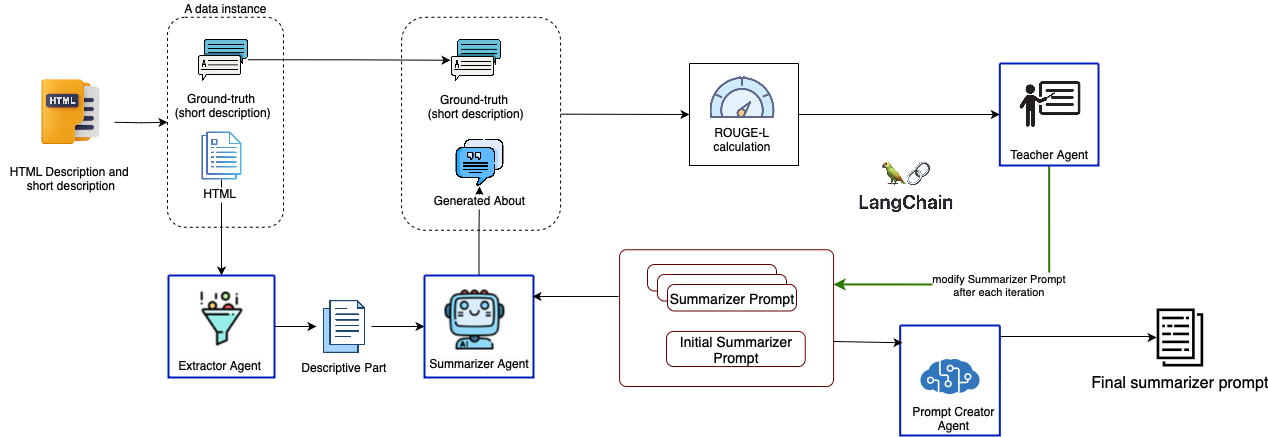}} \\ 
		\subfigure[Evaluation Pipeline]{\label{fig:eval_pipeline}\includegraphics[width=0.88\textwidth]{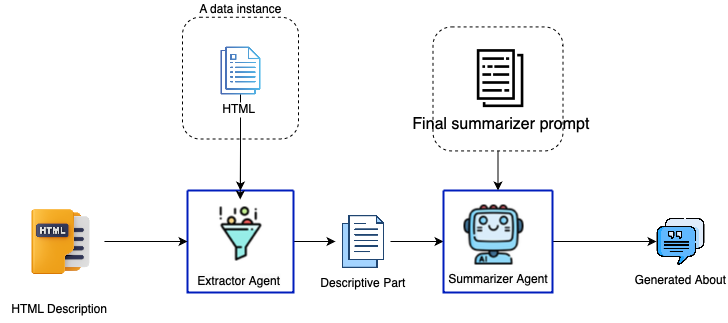}} 
	\end{tabular}
	\caption{Pipelines used in \ME.} 
	\label{fig:pipeline_combined}
\end{figure*}

The pipeline shown in Fig.~\ref{fig:opt_pipeline} involves four specialized LLM-powered agents: \EA, \SA, \TA, and \CA. These agents collaborate through prompt-driven communication, and their outputs are evaluated using ROUGE metrics to iteratively improve the quality of produced summaries as presented below.




\smallskip
\noindent
\textbf{\EA.} This agent is responsible for isolating the most informative parts of a raw description that 
can be in different formats, \eg HTML and Markdown. An application description is typically structured text document that may contain sections such as 
\textit{introduction}, \textit{key features}, \textit{advantages}, \textit{detailed use cases}, \textit{setup instructions}, \textit{compatibility}. 
%
\EA filters out irrelevant or boilerplate content, focusing only on the segments that explain the core function and value proposition of the application. This filtered version, rather than the entire original document, is then passed on for downstream processing, significantly improving both computational efficiency and focus during training.

\smallskip
\noindent
\textbf{\SA.} Given the refined description, this agent generates a concise and meaningful ``About'' statement for the app. Starting from an initial prompt, it iteratively produces summaries that are then improved over multiple rounds using updated prompts generated by \TA. By means of a dedicated prompt, \SA focuses on capturing the app’s primary goal, context, and differentiating features, while excluding peripheral or overly detailed information. The final result is a short and clear description that accurately represents the essence of the application.

\smallskip
\noindent
\textbf{\TA.} It evaluates the quality of each generated summary using the following four elements: \emph{(1)} the current summarization prompt; \emph{(2)} the generated short description; \emph{(3)} the ground-truth target description; and \emph{(4)} the computed ROUGE-L scores. Based on these inputs, it produces an improved prompt for \SA. This optimization follows a multi-step reasoning structure designed to pinpoint mismatches and refine instructions accordingly.

\smallskip
\noindent
\textbf{\CA.} After multiple training iterations across diverse samples, this agent aggregates and generalizes the refined prompts into a single, robust instruction set. It extracts recurring patterns and conditional logic to form a unified prompt suitable for inference. This final prompt helps \SA adapt to varying input description styles and content structures, ensuring consistent output quality.


\smallskip
\ME has been designed so as to allow for a flexible integration of different LLMs. In particular, for this implementation, five LLMs are considered as the engine, including: \texttt{GPT-4o}, \texttt{GPT-3.5-turbo}, \texttt{Gemma}, \texttt{Mistral}, and \texttt{Llama}. For the configuration with \texttt{GPT}, to optimize resource usage, \EA and \SA are implemented using the lightweight \texttt{GPT-4o-mini} version, which offers an efficient balance of performance and cost. In contrast, \TA and \CA leverage the more capable \Fo model to perform higher-level evaluation and reasoning. This hierarchical configuration enables the system to scale effectively while maintaining strong overall performance.



As illustrated in Figure~\ref{fig:opt_pipeline}, the workflow of the agents unfolds as follows:

\begin{enumerate}
	\item A 
	description is provided to \EA, which generates a concise version of the text to be used as input by  \SA.
	
	\mybox{\textbf{\small{\EA Prompt}}}{gray!10}{gray!10}{\small{Your task is to shorten and extract only the introduction and description information from an app. You are given the following description for an app:
			
			$<$Description$>$ \\
			readme\_text \\
			$<$/Description$>$
			
			\# Steps \\
			- **Identify the structure of the app's description**: The app's description is a structure text file that might contains many sections such as introduction, description, key features, advantages, detailed use case, setup instructions, compatibility,... \\
			
			- **Remove all sections that are not relevant to the app's description**: Irrelevant sections might include technical guidance (installing/running/specification... instruction), compatibility, troubleshooting,... \\
			
			- **Remove all unnecessary links/tags**: Identify all links/tags that DO NOT contribute to the description of the app. You must remove all of these reference links and tags. \\
			
			- **Return only text that is relevant to the description of the app**: The output should only contains the text that is relevant to the introduction/description of the app, including the app name/title, app feature description/purpose statement/overview. DO NOT include any output identifications such as: "Here's the ..." or "Extracted App's description:"
			"""
	}}
	
	\item Optimization Loop:
	\begin{enumerate}
		\item \SA receives the extracted text and produces a summary based on its current prompt, shown below.

		\mybox{\textbf{\small{\SA Prompt}}}{gray!10}{gray!10}{\small{Your task is to shorten and extract only the introduction and description information from an app. You are given the following description for an app: \\
				
				``''" \\
				Summarize the following extracted text from an app's description into a short term/phrase introducing the app: \\
				$<$EXTRACTED\_APP'S\_DESCRIPTION$>$ \\
				\$extracted\_text \\
				$<$/EXTRACTED\_APP'S\_DESCRIPTION$>$ \\
				
				The output should include only a short term/phrase introducing the app. \\
				``''"
		}}

		\item A ROUGE-L score is then calculated by comparing the generated summary with the ground truth summary. If the score is equal to or exceeds a predefined threshold, the optimization loop is interrupted, and the current prompt used by \SA is saved as a candidate prompt.
		
		\item If the ROUGE-L score is below the threshold, \TA receives the inputs from \EA and \SA along with the ROUGE-L score. It uses this information to generate a new prompt, which is then passed to \SA for the next iteration. The prompt used by \TA is as follows.

		\mybox{\textbf{\small{\TA Prompt}}}{gray!10}{gray!10}{\small{You are a professional Prompt Engineer. You are working on a system using a Large Language Model (LLM) to help developers automatically generate a short Description term/phrase contain key concept/idea from an extracted text of the description of an app. Your task is to modify and improve the current prompt of the LLM based on the result of testing on a data include a description and a ground truth description. \\
				
				\# Steps: \\
				- **Analyze the data for testing**: Analyze the following data include an extracted text from a description and a ground truth description from an app: \\
				$<$EXTRACTED\_TEXT$>$ \\
				\$extracted\_text \\
				$<$/EXTRACTED\_TEXT$>$ \\
				
				$<$GROUND\_TRUTH DESCRIPTION$>$ \\
				\$description \\
				$<$/GROUND\_TRUTH DESCRIPTION$>$ \\
				- **Review the current result**: Review the generated description using the extracted text its ROUGE score on the ground truth description to identify improvements that could be made: \\
				$<$GENERATED\_DESCRIPTION$>$ \\
				\$generated\_about \\
				$<$/GENERATED\_DESCRIPTION$>$ \\
				$<$ROUGE\_SCORE$>$ \\
				\$rouge\_score \\
				$<$/ROUGE\_SCORE$>$ \\
		}}

		\item The loop continues until one of the following three conditions is satisfied: \emph{(i)} the ROUGE-L score meets or exceeds the threshold; \emph{(ii)} the maximum number of predefined iterations is reached; or \emph{(iii)} the ROUGE-L score stops improving or begins to decline.
	\end{enumerate}
	
	\item Once the loop ends, a new HTML description is passed through the same pipeline to generate another candidate prompt.
	
	\item After all HTML descriptions have been processed, \CA collects all candidate prompts that achieved a ROUGE-L score above the defined threshold and aggregates them into a single, final prompt. The agent performs its activities by means of the following prompt.

	\mybox{\textbf{\small{\CA Prompt}}}{gray!10}{gray!10}{\small{You are a professional Prompt Engineer. You are working on a system using a Large Language Model (LLM) to help developers automatically generate a short Description term/phrase contain key concept/idea from an extracted text of the description of an app. Your task is to combine several candidate prompts for the LLM into a final prompt. \\
			
			\# Steps: \\
			- **Review all candidate prompts**: Analyze the following prompts to identify common parts to be included in the final prompt and also includes specific details or conditional key points from these prompts to be included in the final prompt \\

			$<$CANDIDATE\_PROMPTS$>$ \\
			\$summarizer\_list \\
			$<$/CANDIDATE\_PROMPTS$>$ \\
			- **Generate a final prompt**: Based on the common parts and conditional key points, generate a final prompt for the LLM. \\
			
			\# Output Format:\\
			Do not include any reasoning/explanation like "Based on the result of the above review:", "Here's the", ... or any output identifiers like "Prompt:", "New Prompt", ... The output should only include a string representing the prompt for the LLM
			"""			
	}}

\end{enumerate}

\subsection{Orchestration of Agents}

\ME operates as a collaborative pipeline in which multiple LLM-based agents interact toward a common objective. The orchestration of these agents is structured around three main phases: \textit{Agent Communication}, \textit{Iterative Refinement}, and \textit{Prompt Consolidation}, explained as follows.

\begin{itemize}
	\item \textit{Agent Communication}. Each agent functions not in isolation but through controlled interaction with its environment and its peers. To facilitate this, we utilize the \texttt{LangChain} framework as the communication backbone. In addition, outputs are structured in a machine-readable format, enabling agents to reliably pass information to one another. This design ensures consistency and seamless cooperation throughout the pipeline.
	
	\item \textit{Iterative Refinement}. The process begins with \SA receiving a filtered version of the application's HTML description from \EA. Using the current summarization prompt, it generates a candidate ``About'' summary. The result is evaluated using the ROUGE-L metric, which serves as a primary indicator of semantic and structural alignment with the ground-truth summary. This feedback, along with the current prompt and outputs, is passed to \TA, which formulates an improved prompt for the next round.
	
	Unlike traditional fixed-iteration loops as proposed in our previous work \citep{10.1145/3696630.3728511}, in this paper we implement a \textit{dynamic stopping criterion} to optimize the training efficiency as follows:
	
	\begin{enumerate}
		\item Each sample starts with a baseline allowance of 15 iterations.
		\item If the ROUGE-L score stagnates across 3 consecutive iterations or decreases in 2 iterations (with each drop also counted as stagnation), the process halts early for that sample.
		\item If ROUGE-L reaches a predefined threshold (0.7), then the process stops.
	\end{enumerate}
	
	We expect this adaptive mechanism prevents unnecessary computation on low-potential samples, while still offering more iteration budget to promising ones. In addition, we examine the impact of this dynamic iteration strategy in comparison to a non-dynamic (fixed-loop) setting to evaluate gains in performance and efficiency.
	
	\item \textit{Prompt Consolidation}. After the optimization process completes, we gather a pool of prompt versions, each tailored to a specific training instance. These prompts are analyzed by \CA, which extracts common structures and key conditional instructions to synthesize a generalized summarization prompt. This final prompt is then used during inference to ensure that \SA performs effectively across a diverse range of HTML description inputs, maintaining both coherence and relevance in the generated ``About'' summaries.
	
\end{itemize}

\subsection{Evaluation Pipeline}

Once the optimized prompt has been obtained from the Optimization Pipeline, the evaluation phase can then be initiated. As illustrated in Figure~\ref{fig:eval_pipeline}, the Evaluation Pipeline is comparatively simple and consists of only two agents, \ie \EA and \SA. The former uses the same prompt and performs the same function as in the optimization phase, processing the HTML description to extract a concise description. The latter, however, uses the final optimized prompt produced in the previous phase for all summarization. After the summarization, ROUGE-1, ROUGE-2, ROUGE-L and cosine similarity were collected to evaluate the quality of the generated summaries against their respective ground truth references. These metrics are collected from the evaluation of several HTML descriptions so that the optimization task could be assessed.

	\section{Evaluation}	
	\label{sec:Evaluation}
	This section presents the empirical study conducted to evaluate the proposed approach's performance. We introduce the 
research questions in Section~\ref{sec:ResearchQuestions}, 
and describe the datasets in Section~\ref{sec:Dataset}. 


\subsection{Research Questions} \label{sec:ResearchQuestions}

To evaluate \ME we answer the following research questions, and compare it with baselines.

\begin{itemize}

	
	\item \rqfirst~\\
	We investigate if the use of a multi-agent architecture is really needed, given that a single agent might already be sufficient to get a decent recommendation. For this RQ, we consider \texttt{Mixtral-8x7B-Instruct-v0.1}, \texttt{Llama-2-7b-hf}, \texttt{GPT-4o}, and \texttt{Gemma-2-2b-it} as baselines for comparison as they have been widely used in summarization tasks. 
	
	\item \rqsecond~\\This RQ evaluates whether the newly proposed dynamic iteration strategy is beneficial to the generation of short summaries, \ie effective and/or efficient, compared to the static one conceived in our previous work  \citep{10.1145/3696630.3728511}. 
	\item \rqthird~The previous version of \ME relies on \texttt{Original LLM set} including \texttt{GPT-4o} and \texttt{GPT-4o-mini} 
	as the only engine for agents. In this paper, we further extended the architecture to sustain other LLMs, including \texttt{Mistral-7B-Ins\-truct-v0.3}, \texttt{Llama-3.2-3B-Instruct}, and \texttt{Gemma-2-2b-it}, to validate the extensibility of \ME.
	
	
\end{itemize}

\subsection{Datasets} \label{sec:Dataset}

In this section, we explain in detail the process conducted to curate, clean, and restructure the datasets used in the evaluation.

\subsubsection{Data Curation}

To assess the effectiveness of \ME in a domain beyond traditional software documentation, we curated a novel dataset of mobile applications derived from the AndroZoo\footnote{\url{https://androzoo.uni.lu/}} repository~\citep{Allix:2016:ACM:2901739.2903508}, a large-scale archive comprising over 25 million Android applications mined from various sources.
We initially selected a representative sample of 20{,}000 applications from the 5GB AndroZoo summary metadata file, filtering for apps that \textit{(i)} were sourced from the Google Play Store; and \textit{(ii)} had a release date after January 1st, 2020 (we consider apps in the most recent 5 years). 
Application metadata was retrieved programmatically through the AndroZoo API, which provides multiple versions for each application along with associated attributes (e.g., \texttt{name}, \texttt{description\_short}, \texttt{description\_html}, \texttt{version\_\-string}).

The following filtering criteria were applied to refine the corpus:

\begin{itemize}
	\item Apps with only one version available were excluded.
	\item For multi-version apps, only the metadata of the most recent version was retained.
	\item Apps without a star rating (\ie no user reviews) were discarded.
	\item Using the \texttt{langdetect}\footnote{\url{https://github.com/Mimino666/langdetect}} library for language identification, we excluded apps whose \texttt{description\_html} or \texttt{description\_short} fields are not written in English.
\end{itemize}

After filtering, the final dataset consisted of 2{,}980 mobile applications. For each application, we used the \texttt{description\_html} field as input and the corresponding \texttt{description\_short} as the ground-truth summary. 
Table~\ref{tab:star-distribution} reports the number of applications falling into predefined rating intervals.

\begin{table}[h]
	\centering
	\caption{Distribution of applications by average star rating.}
	\label{tab:star-distribution}
	\begin{tabular}{|l|c|} \hline
		\textbf{Star Rating Range} & \textbf{Number of Applications} \\ \hline
		0--1     & 0   \\ \hline
		1--2     & 68  \\ \hline
		2--3     & 361 \\ \hline
		3--4     & 854 \\ \hline
		4--5     & 1,696 \\ \hline

	\end{tabular}
\end{table}

As shown in the table, the majority of the apps in the dataset have a rating above 3.0, with over half being rated between 4 and 5. This distribution reflects a natural skew toward well-rated apps, which are also more likely to have informative and well-structured descriptions. Apps without any rating were excluded as part of our data quality filtering process, 
and the statistics are depicted in Table~\ref{tab:dataset-stats}. 

\begin{table}[h]
	\centering
	\caption{Summary statistics of the mobile application dataset (token-level estimates).}
	\label{tab:dataset-stats}
	\begin{tabular}{|l|r|r|} \hline
		\textbf{Metric} &  \texttt{description\_html} & \texttt{description\_short} \\ \hline
		Average Length       & 289 tokens  & 10 tokens \\ \hline
		Median Length        & 244 tokens  & 11 tokens \\ \hline
		Minimum Length       & 2 tokens    & 1 token \\ \hline
		Maximum Length       & 798 tokens  & 17 tokens \\ \hline
		Standard Deviation   & 198 tokens  & 3 tokens \\ \hline
	\end{tabular}
\end{table}

A distributional analysis reveals that input texts exhibit high variance and right-skewness, consistent with the verbose and heterogeneous nature of mobile app descriptions. In contrast, the ground-truth summaries are concise, with low variance and a narrow range, confirming the suitability of this dataset for evaluating the summarization capabilities of \ME.

\subsubsection{Data Preprocessing} 

\begin{figure}[ht]
	\centering
	\includegraphics[width=\linewidth]{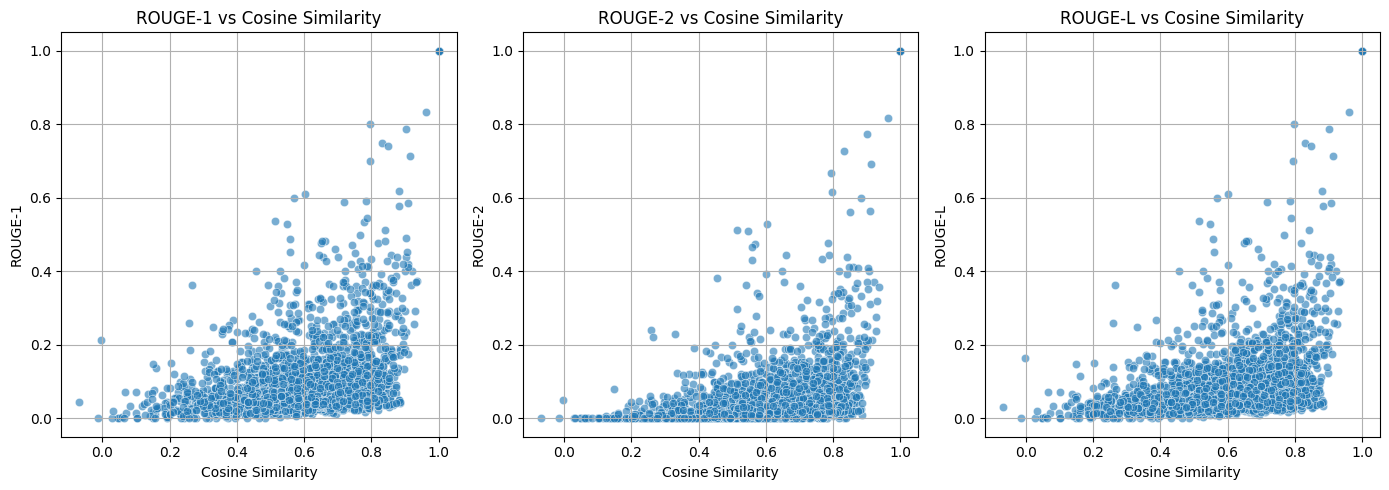}
	\caption{The relationship between Cosine Similarity and ROUGE scores.} 
	\label{fig:rouge_cosine_scatter}
\end{figure}

Data quality plays a crucial role in the fine tuning and inference phases. To assess the quality of these pairs, we conducted an exploratory data analysis (EDA) by computing ROUGE scores (ROUGE-1, ROUGE-2, ROUGE-L) and Cosine Similarity between \texttt{description\_html} and \texttt{description\_short}, and the scores are shown in Figure~\ref{fig:rouge_cosine_scatter}. 

As seen in the figure, there is a clear positive correlation between ROUGE and Cosine Similarity scores. In other words, a higher semantic similarity 
tends to coincide with a denser n-gram overlap. 
This observation allows us to come up with the following 
data filtering strategy.

\begin{itemize}
	\item \textbf{Training set}. Given the goal of building a system that learns effectively from a small amount of high-quality data, we selected training samples with 0.85 $\leq$ Cosine Similarity $<$ 1.0.  
	This range ensures that each training pair is highly relevant yet not identical, thereby reducing noise and enhancing learning efficiency.
	\item \textbf{Test set}. It was constructed from the remaining samples not used in training. The following filtering rule was applied:  
	
	\begin{center}
		\textit{(ROUGE-1~$\geq$~0.1) or (ROUGE-2~$\geq$~0.1) or (ROUGE-L~$\geq$~0.1)} \textbf{and} \textit{Cosine Similarity~$\geq$~0.4}
	\end{center}
	
\end{itemize}

These thresholds were selected based on empirical inspection of samples at various cutoff points, ensuring a reasonable level of summary relevance while maintaining diversity. This filtered set contains over 800 samples, and it is used to derive the following final sets for the evaluation. 

\begin{itemize}
	\item \textbf{D$_1$}: 400 randomly sampled test cases.
	\item \textbf{D$_2$}: 600 randomly sampled test cases.
	\item \textbf{D$_3$}: Top 200 samples with the highest Cosine Similarity scores.
	\item \textbf{D$_4$}: Top 200 samples with the highest ROUGE-L scores.
\end{itemize}

These subsets enable us to evaluate the performance of \ME across both general and high-quality cases, facilitating robust and diverse performance analysis. The next section elaborates on 
the results obtained from the empirical evaluation.

	\section{Experimental Results}	
	\label{sec:Results}

We report and analyze the experimental results by answering the 
research questions introduced in Section~\ref{sec:ResearchQuestions}.

\subsection{\rqfirst} 

In our previous work \citep{10.1145/3696630.3728511}, \ME was compared with various studies, and the 
results demonstrated that it consistently outperformed the baselines in generating repository descriptions from \GH \RM files. 
Recent studies have evaluated the performance of different LLMs on summarization tasks across various domains~\citep{Aly2025,takeshita-etal-2025-irsum,10.1145/3677389.3702588,10986332}. These studies highlight that models such as \texttt{Mixtral-8x7B-Instruct-v0.1} and \texttt{Llama-2-7b-hf} achieve competitive or superior results, especially in zero-shot settings. To further deepen 
the comparison with single-agent LLMs, we extended our evaluation to include two widely adopted general-purpose LLMs that are frequently used in summarization tasks including \texttt{GPT-4o} and \texttt{Gemma-2-2b-it}.

\begin{landscape}
	\begin{figure*}[t!]
		\centering    
		\begin{tabular}{c}		
			\subfigure[\textbf{D$_1$}]{\label{fig:RQ1-ROUGE-D1}\includegraphics[width=0.95\linewidth]{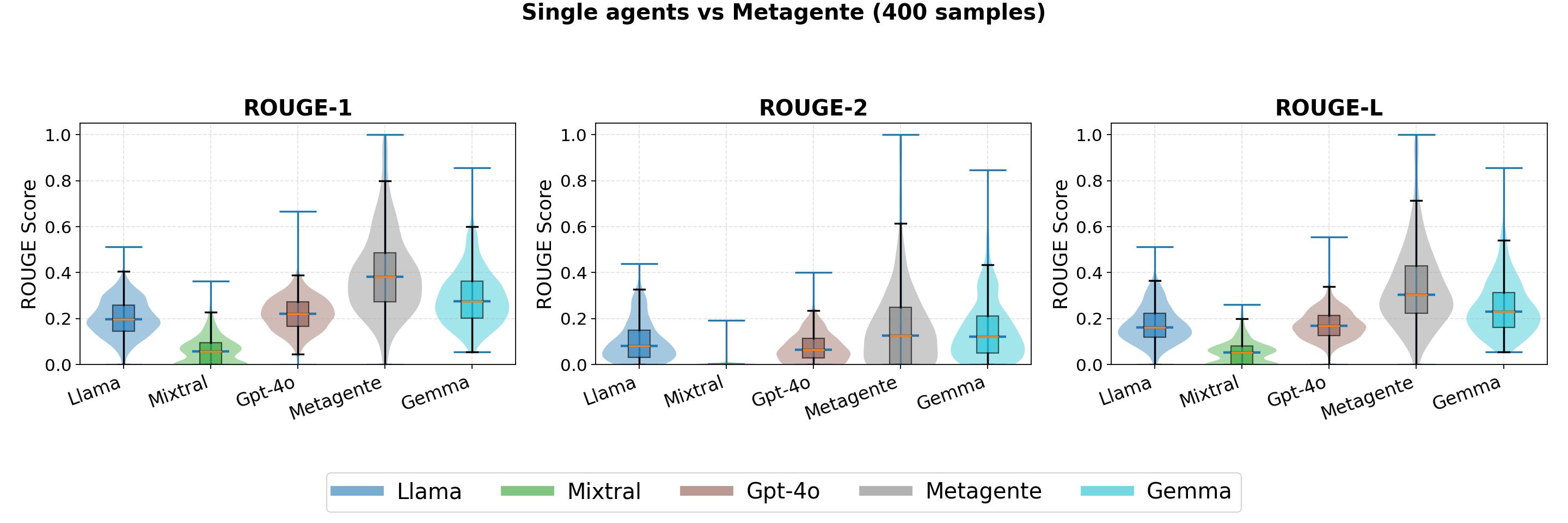}} \\ 
			\subfigure[\textbf{D$_2$}]{\label{fig:RQ1-ROUGE-D2}\includegraphics[width=0.95\linewidth]{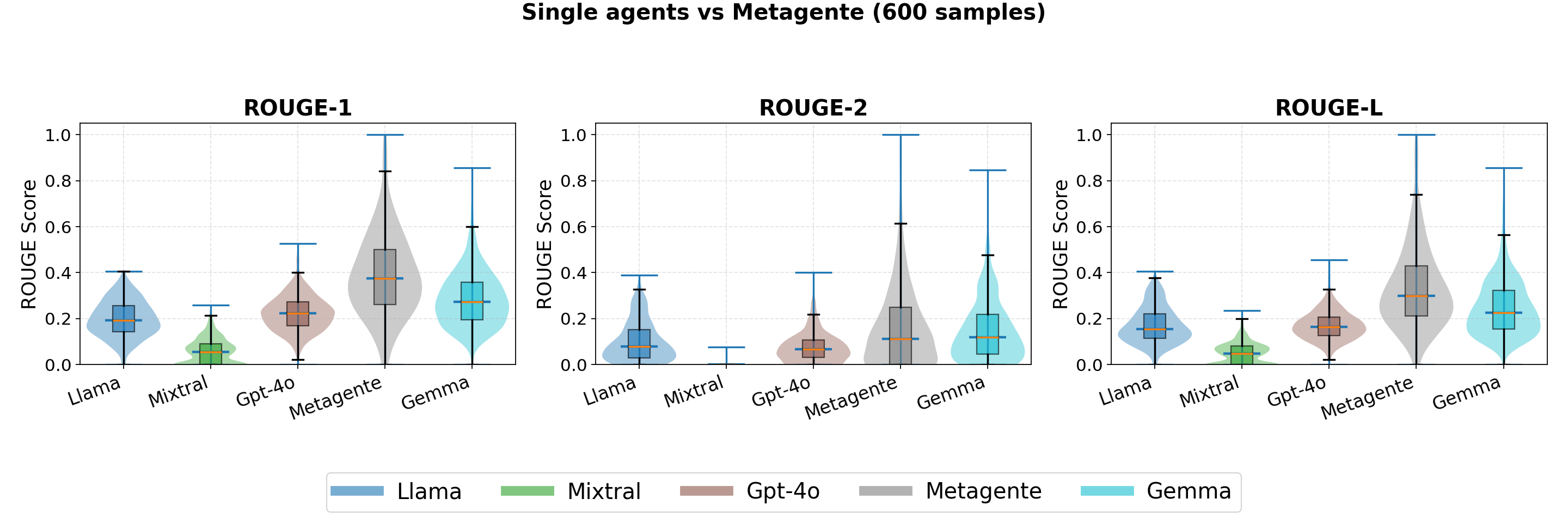}} 
		\end{tabular}
		\caption{Comparison between \ME and LLM-based single agents on the \textbf{D$_1$} and \textbf{D$_2$} datasets.} 
		\label{fig:RQ1-ROUGE-D1D2}
	\end{figure*}
	
\end{landscape}

We evaluate \ME against the baselines using two 
datasets, \ie \textbf{D$_1$} and \textbf{D$_2$} and the results are shown in Figure~\ref{fig:RQ1-ROUGE-D1D2}. 
Overall, the generated 
summaries exhibit higher or comparable semantic similarity (cosine) and competitive ROUGE scores compared to ground-truth summaries. As shown in Figure~\ref{fig:RQ1-ROUGE-D1D2}, it is evident that both the baselines, \ie \texttt{Mixtral-8x7B-Instruct-v0.1} and \texttt{Llama-2-7b-hf}, achieve a lower and less stable performance on the two datasets. An extremely low recommendation quality is seen by \texttt{Mixtral-8x7B-Ins\-truct-v0.1}, \ie most of the ROUGE scores are smaller than 0.15.  Their outputs also show a greater variance--some summaries are overly generic, while others omit key aspects of the original descriptions. This underscores the baselines' limitations when used for summarization without iterative refinement or agent collaboration.

The figure shows that \ME consistently outperforms both baselines across all the evaluation metrics, \ie 
it delivers more consistent and higher-quality results across both random subsets. The advantage is especially clear in ROUGE-L, which captures structural and semantic similarity with the reference summaries. In particular, by the \textbf{D$_1$} dataset (see Figure~\ref{fig:RQ1-ROUGE-D1}), most of the ROUGE-L scores obtained by \ME range from 0.3 to 1.0. Meanwhile by \texttt{Mixtral-8x7B-Instruct-v0.1} and \texttt{Llama-2-7b-hf}, the corresponding scores are much smaller, \eg less than 0.2. The same trend is witnessed with \textbf{D$_2$} (see Figure~\ref{fig:RQ1-ROUGE-D2}), in which the ROUGE scores obtained by \ME are always superior to those of \texttt{Mixtral-8x7B-Instruct-v0.1} and \texttt{Llama-2-7b-hf}.   

This suggests that \ME\ generates summaries that are not only concise but also more faithful to the original content, reflecting both the overall purpose and the unique details of the applications described. These improvements point to the value of the multi-agent optimization loop in producing summaries that are easier to read and more informative. The stability of these improvements across metrics and datasets provides strong evidence that the Teacher–Student refinement loop and prompt consolidation strategy not only improve accuracy but also enhance robustness and generalizability in diverse summarization settings.

\noindent\fbox{\begin{minipage}{0.98\columnwidth}
		\paragraph{\textbf{Answer to RQ$_1$:}} \ME consistently outperforms various single LLM-based agents in terms of all the ROUGE scores, 
		demonstrating the benefit of collaboratively joint LLM-based agents. 
\end{minipage}}

\begin{landscape}
	
	\begin{figure*}[t!]
		\centering    
		\begin{tabular}{c}		
			\subfigure[Comparison of generated summary with ground-truth \texttt{description\_short}]{\label{fig:RQ2-GroundTruth_DescriptionShort}\includegraphics[width=0.95\linewidth]{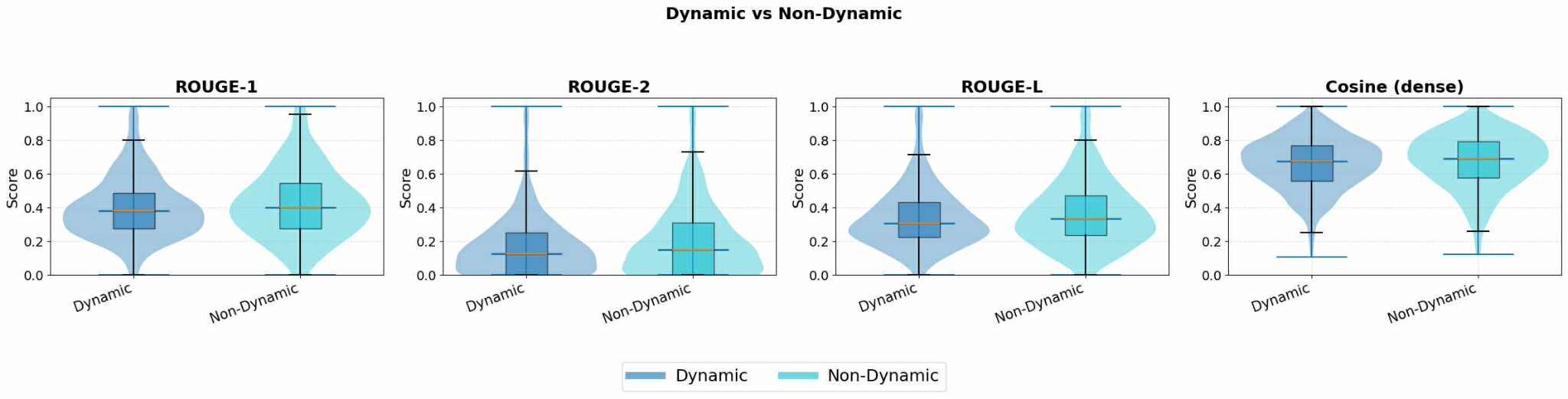}} \\ 
			\subfigure[Comparison of generated summary with ground-truth \texttt{description\_html}]{\label{fig:RQ2-GroundTruth_README}\includegraphics[width=0.95\linewidth]{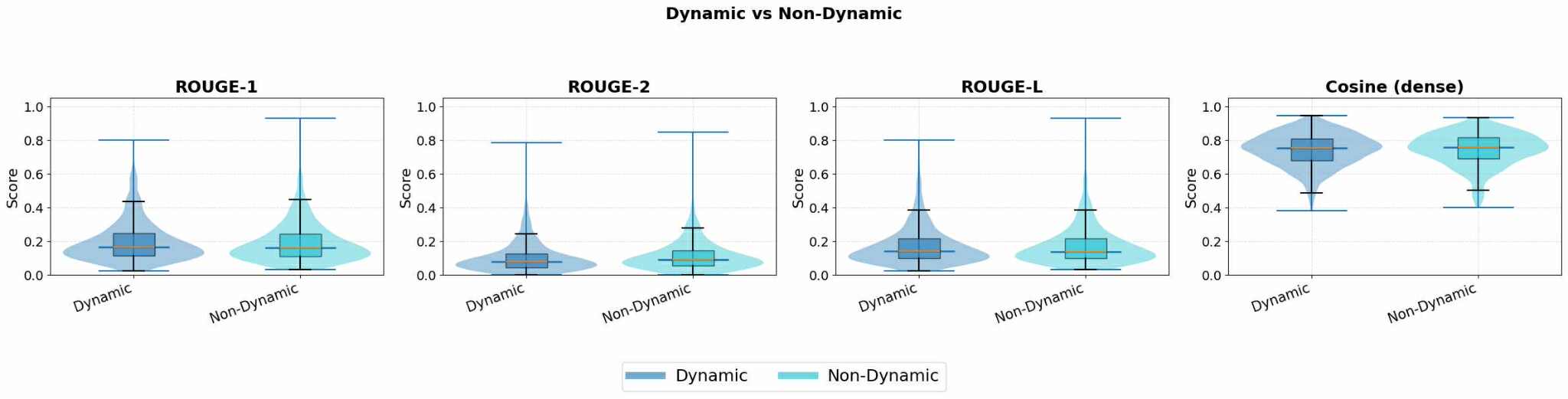}} 
		\end{tabular}
		\caption{Dynamic vs. Non-Dynamic strategies.} 
		\label{fig:RQ2-Comparison}
	\end{figure*}
	
\end{landscape}


\subsection{\rqsecond} 


We conducted a comparative analysis between the dynamic and non-dynamic prompting strategies using ROUGE scores and cosine similarity. The results are depicted in Figure~\ref{fig:RQ2-Comparison}.

Quantitatively, the non-dynamic approach shows slightly higher averages in ROUGE-2 and ROUGE-L, suggesting stronger n-gram overlap with reference summaries. In contrast, the dynamic method performs on par in ROUGE-1 and demonstrates stronger semantic alignment in cosine similarity, highlighting its ability to capture meaning beyond surface-level overlap.

\begin{figure*}[h!]
	\centering
	\begin{tabular}{c c}	
		\subfigure[Average Flow Duration]{\label{fig:FlowDuration}
			\includegraphics[width=0.45\linewidth]{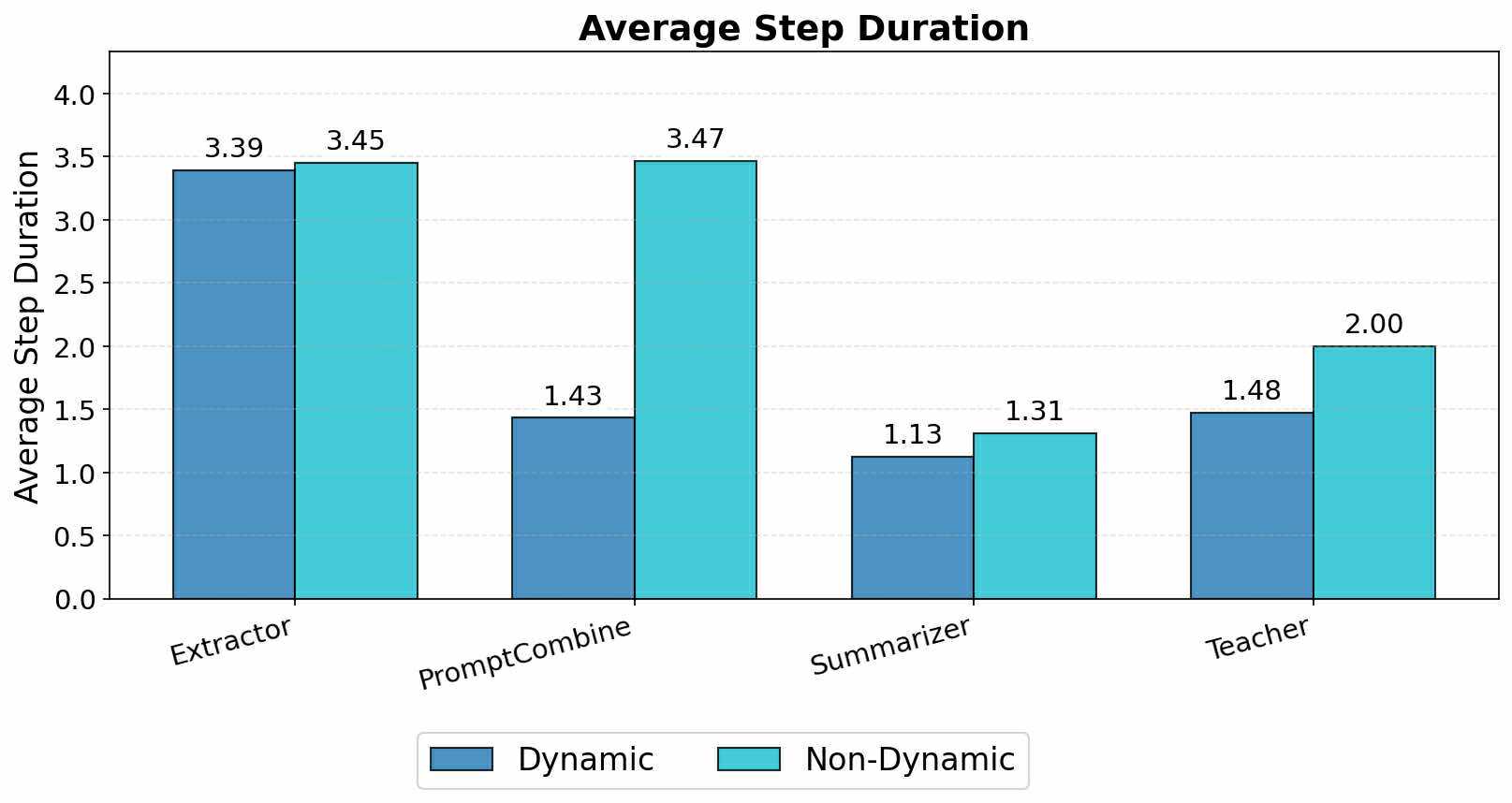}}	&
		\subfigure[Average Step Duration]{\label{fig:StepDuration}
			\includegraphics[width=0.40\linewidth]{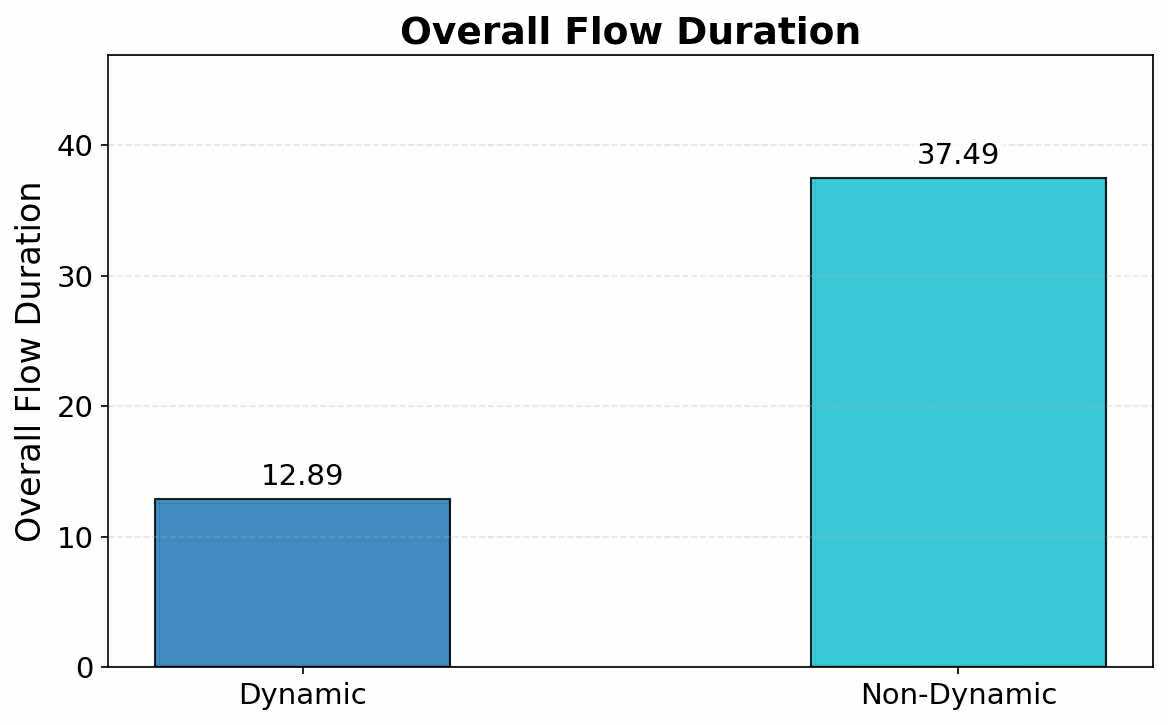}} \\
			
		\subfigure[Number of input tokens]{\label{fig:InputTokens}
			\includegraphics[width=0.45\linewidth]{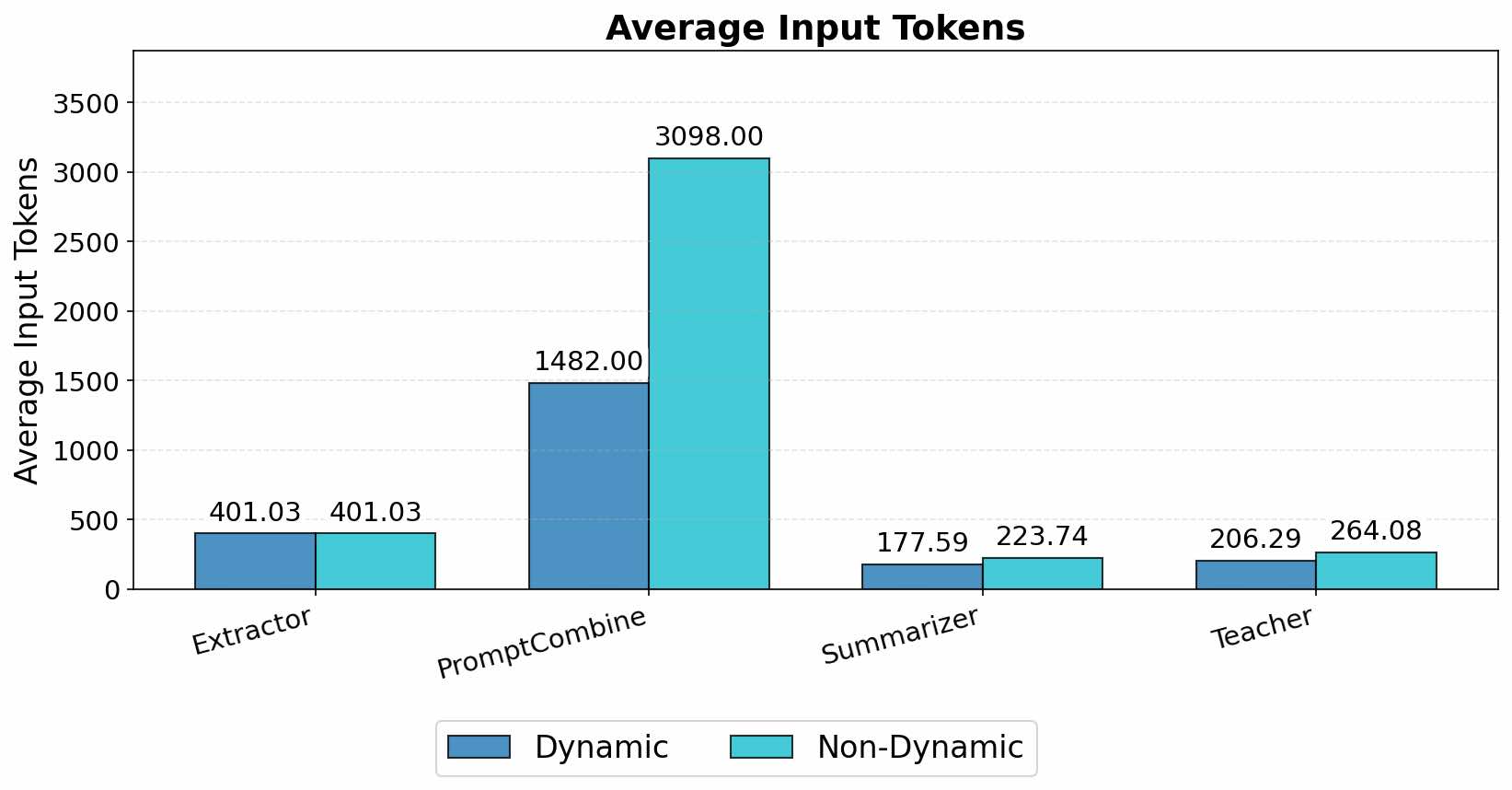}}	&
		\subfigure[Number of output tokens]{\label{fig:OutputTokens}
			\includegraphics[width=0.45\linewidth]{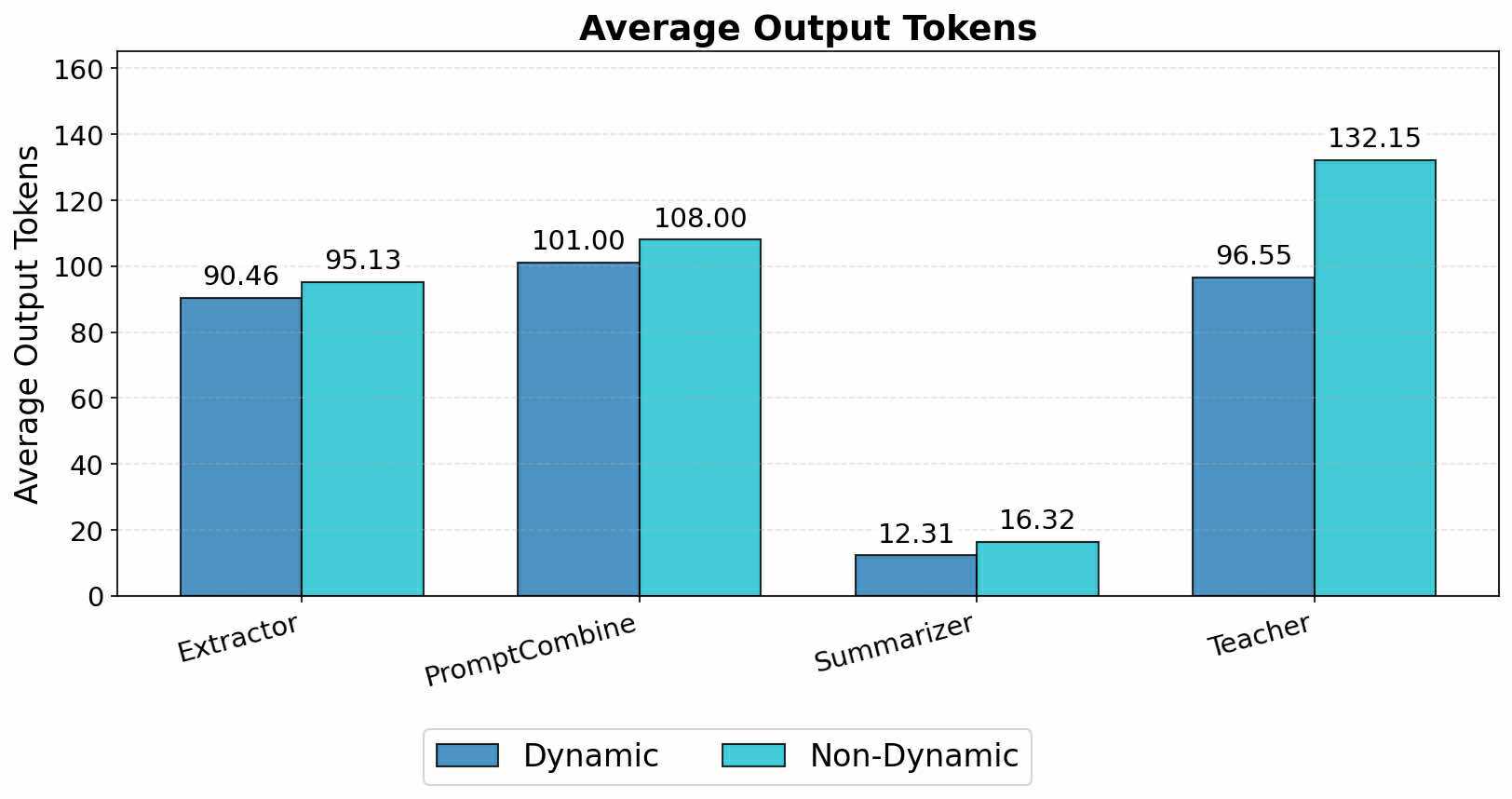}}

	\end{tabular} 
	\caption{Comparison of efficiency between non-dynamic and dynamic strategies.} 	
	\label{fig:Efficiency}
\end{figure*}

Interestingly, despite comparable performance, the dynamic strategy completed training in only 260 iterations, while the non-dynamic method requires 620 iterations to converge. The comparison of efficiency between non-dynamic and dynamic strategies is shown in Figure~\ref{fig:Efficiency}. 

The figure shows that compared to the non-dynamic method, the dynamic one is always efficient with respect to the timing and resources used. This suggests that the dynamic method is significantly more efficient in learning useful prompts, adapting early based on intermediate feedback. 
Figure~\ref{fig:RQ2-Comparison} further highlights that both strategies follow similar trends in ROUGE and cosine similarity distributions. However, the dynamic version tends to produce outputs with higher semantic consistency and broader generalization, whereas the non-dynamic one exhibits sharper peaks in overlap-based metrics.

The evaluation reveals that, despite producing results comparable to the non-dynamic setting, the dynamic iteration strategy requires significantly fewer training steps. This means that adaptive stopping criteria help avoid wasted computation while still capturing meaningful improvements. Such efficiency is especially important in real-world deployments, where computational resources are often limited.  

\vspace{.2cm}
\noindent\fbox{\begin{minipage}{0.98\columnwidth}
		\paragraph{\textbf{Answer to RQ$_2$:}}
		The dynamic method creates summaries 
		comparable to the those generated by the 
		non-dynamic ones. However, the dynamic method needs 
		much fewer training steps (260 vs. 620). 
		This means that it improves efficiency by reducing unnecessary training steps while maintaining comparable quality, making the framework suitable for resource-constrained environments in real-world scenarios.
\end{minipage}}

\subsection{\rqthird} 

To evaluate the compatibility and adaptability of our pipeline across various LLMs, we implemented \ME using multiple models--ranging from lightweight open-source LLMs to commercial APIs--on the same datasets and measured their performance using ROUGE metrics. In particular, the following models are considered in our evaluation: \texttt{Original LLM set (a combination of GPT-4o and GPT-4o-mini)}, \texttt{GPT-3.5-turbo}, \texttt{Gemma-2-2b-it}, \texttt{Mistral-7b-I\-nstruct-v0.3}, and \texttt{Llama-3.2-3B-Instruct}. 

Figure~\ref{fig:ROUGE-D1D2} and Figure \ref{fig:ROUGE-D3D4} depict violin boxplots, comparing the average ROUGE-1, ROUGE-2, and ROUGE-L scores across four evaluation subsets: two random test sets, \ie \textbf{D$_1$} and \textbf{D$_2$} with 400 and 600 samples, respectively, and two top-ranked subsets based on Cosine Similarity and ROUGE-L, \textbf{D$_3$} and \textbf{D$_4$} each with 200 samples.

As shown in the figures, our pipeline demonstrates compatibility with all tested LLMs 
although performance varies--with the original LLM set consistently outperforming others--the system remains operational and effective with all models. The framework remains robust across different LLM backbones, indicating strong portability and adaptability in diverse real-world scenarios. Even lighter models such as \texttt{Gemma-2-2b-it} and \texttt{Mistral-7B-Ins\-truct-v0.3} were able to produce summaries meeting minimum ROUGE thresholds, indicating robustness in prompt processing and adaptation.

\begin{landscape}
	
	\begin{figure*}[t!]
		\centering    
		\begin{tabular}{c}		
			\subfigure[\textbf{D$_1$}]{\label{fig:ROUGE-D1}\includegraphics[width=0.95\linewidth]{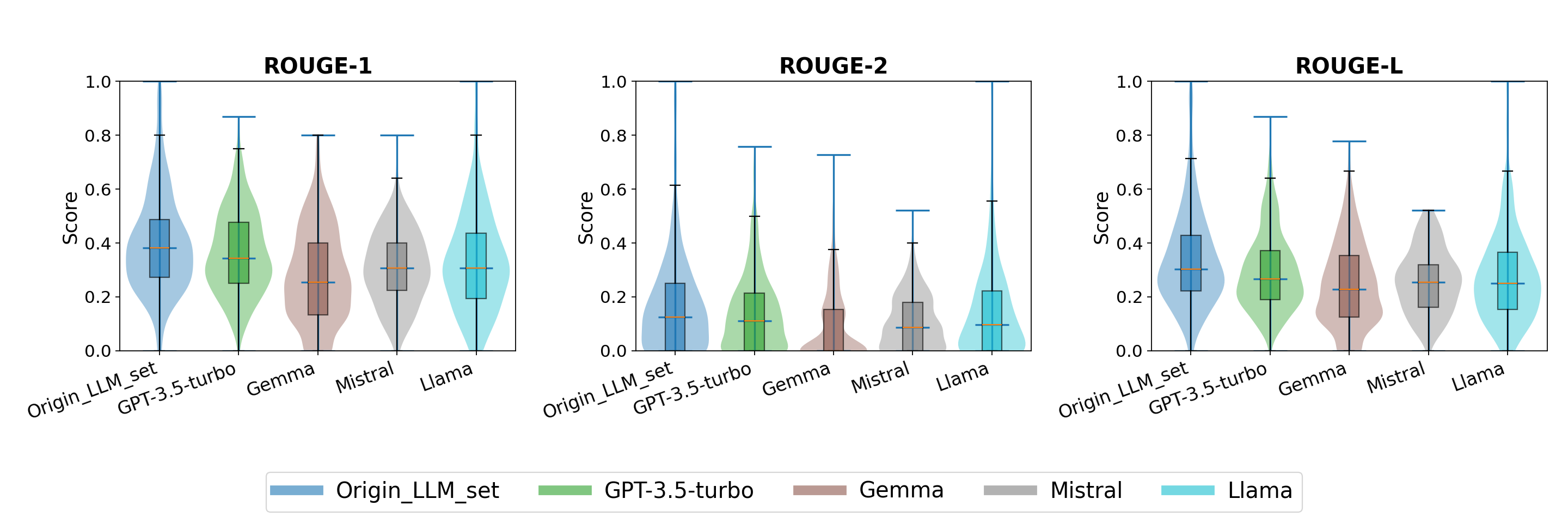}} \\ 
			\subfigure[\textbf{D$_2$}]{\label{fig:ROUGE-D2}\includegraphics[width=0.95\linewidth]{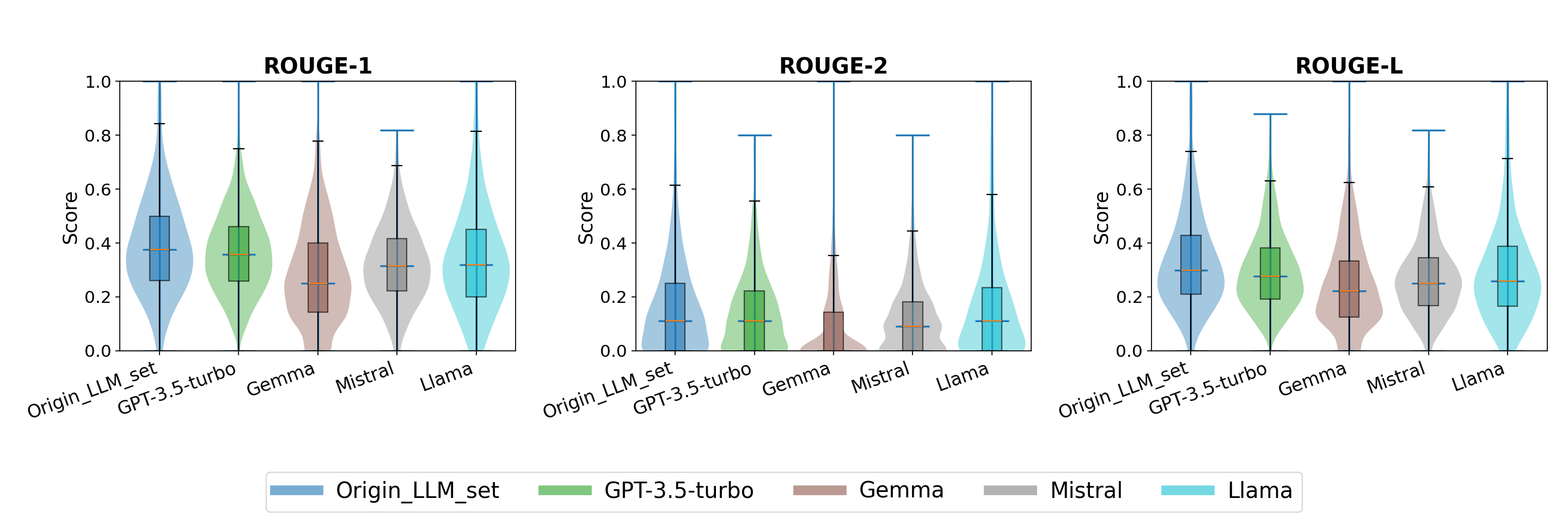}} 
		\end{tabular}
		\caption{Performance comparison on \textbf{D$_1$} and \textbf{D$_2$}.} 
		\label{fig:ROUGE-D1D2}
	\end{figure*}

	\begin{figure*}[t!]
		\centering    
		\begin{tabular}{c}		
			\subfigure[\textbf{D$_3$}]{\label{fig:ROUGE-D3}\includegraphics[width=0.95\linewidth]{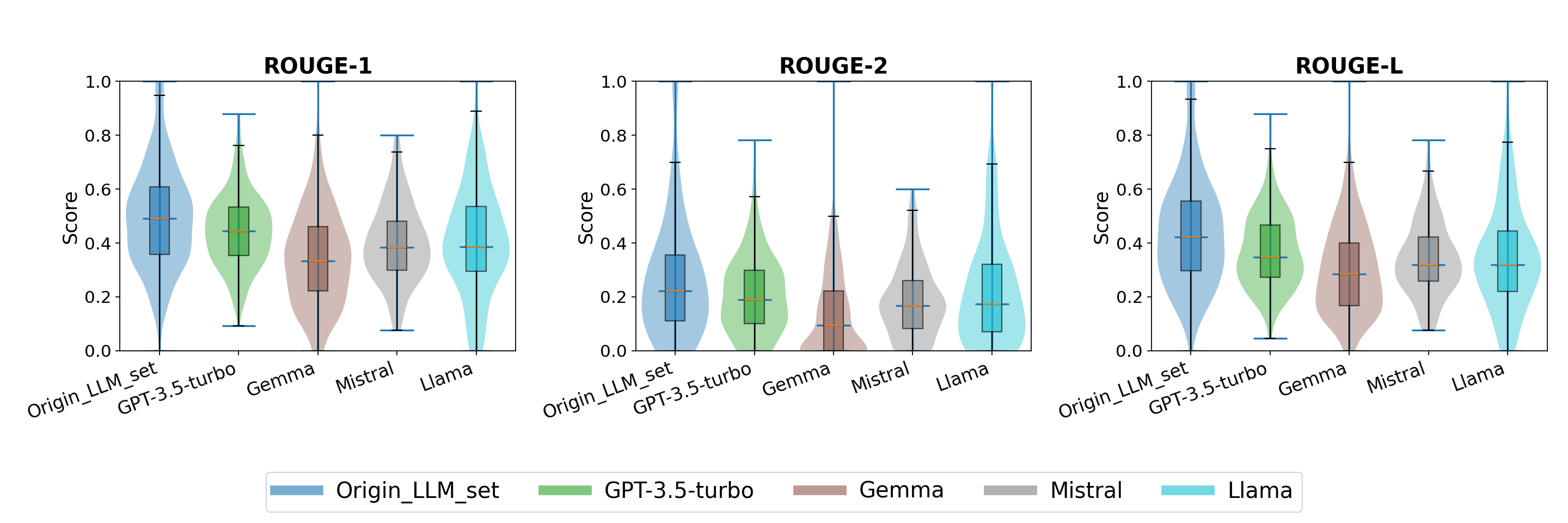}} \\ 
			\subfigure[\textbf{D$_4$}]{\label{fig:ROUGE-D4}\includegraphics[width=0.95\linewidth]{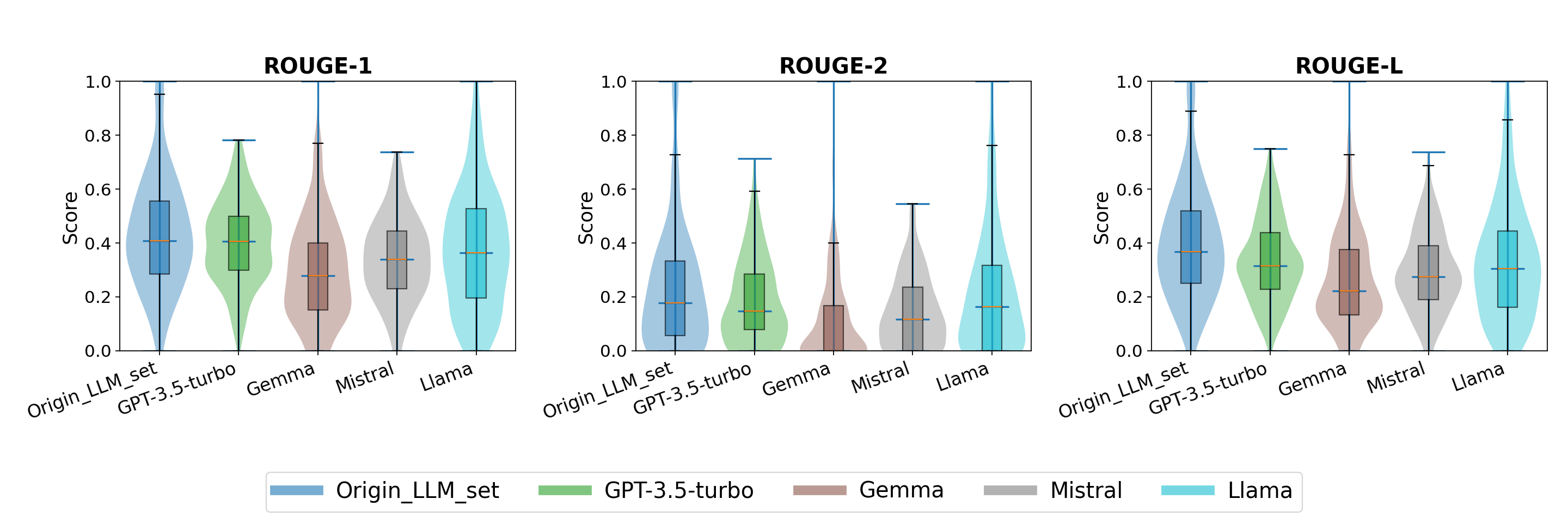}} 
		\end{tabular}
		\caption{Performance comparison on \textbf{D$_3$} and \textbf{D$_4$}.} 
		\label{fig:ROUGE-D3D4}
	\end{figure*}
	
\end{landscape}


In Figure~\ref{fig:ROUGE-D1D2}, concerning the ROUGE scores obtained on \textbf{D$_1$}, we can see that using \texttt{Gemma-2-2b-it}, \ME yields a mediocre performance compared to that when using the other LLMs. \texttt{Mistral-7B-Ins\-truct-v0.3} as the agent is better than \texttt{Gemma-2-2b-it}, still it underperforms the remaining LLMs including  \texttt{Original LLM set},  \texttt{GPT-3.5-turbo} and  \texttt{Llama-3.2-3B-Instruct}. Among others, the \texttt{Original LLM set} with two \texttt{GPT-4o} agents and \texttt{GPT-4o-mini} agents contributes to the best performance to \ME. The same trend is seen on \textbf{D$_2$} (shown in Figure~\ref{fig:ROUGE-D2}). These findings suggest that our pipeline does not depend on a specific backbone model, making it portable and flexible for different deployment scenarios, including low-resource settings.

\begin{figure*}[t!]
	\centering    
	\begin{tabular}{c}		
		\subfigure[ROUGE-1]{\label{fig:avg_ROUGE-1}\includegraphics[width=0.60\textwidth]{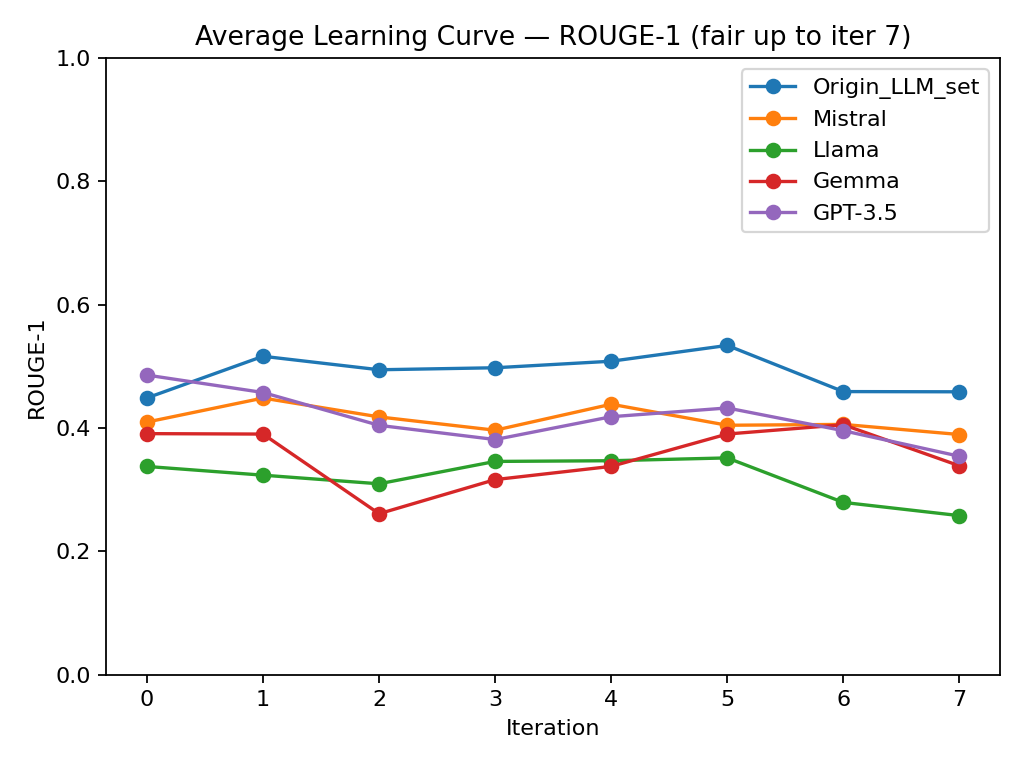}} \\ 
		\subfigure[ROUGE-2]{\label{fig:avg_ROUGE-2}\includegraphics[width=0.60\textwidth]{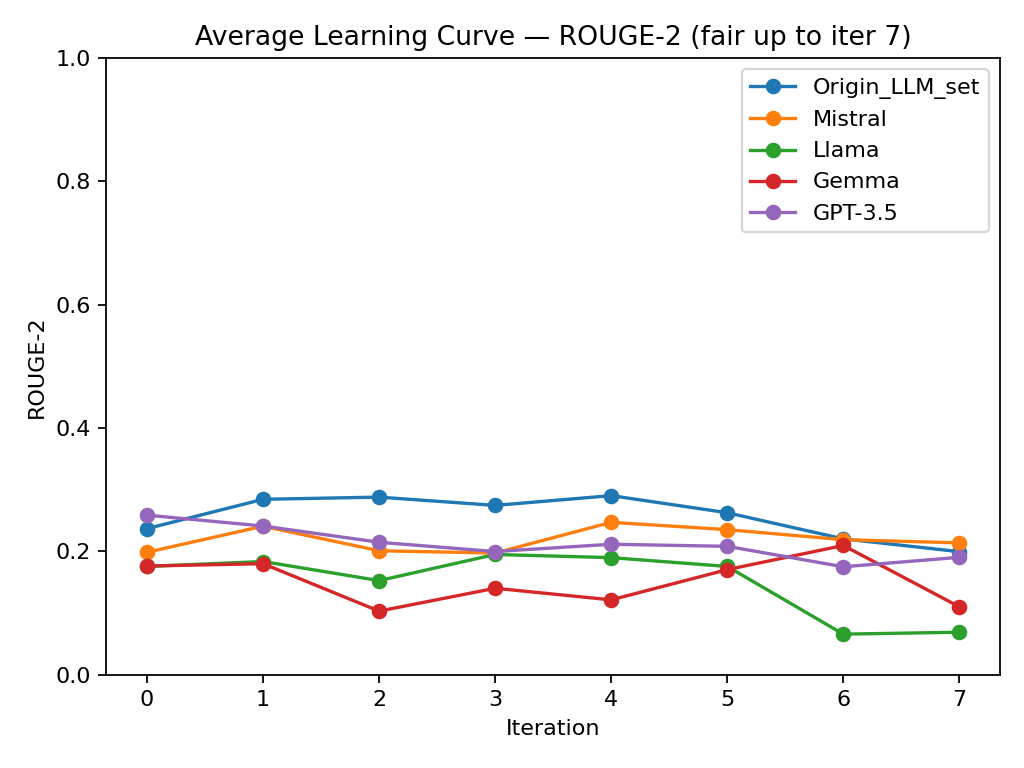}} \\
		\subfigure[ROUGE-L]{\label{fig:avg_ROUGE-L}\includegraphics[width=0.60\textwidth]{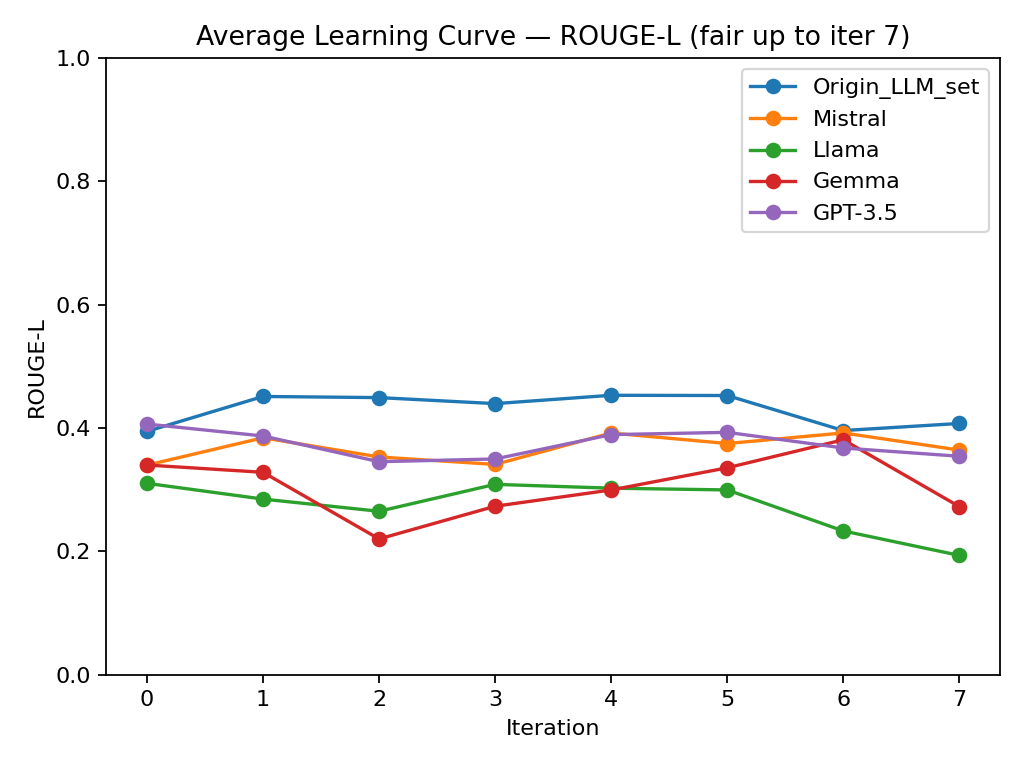}} 
	\end{tabular}
	\caption{Learning curves.} 
	\label{fig:LearningCurves}
\end{figure*}

%
%

We further investigated how the system improves summary quality over time using the proposed Teacher-Student architecture. For this, we keep track of the ROUGE scores across iterations for each model and computed the average learning curves for ROUGE-L, ROUGE-2, and ROUGE-1 respectively, up to iteration 7. The final results are depicted in Figure~\ref{fig:LearningCurves}. 

The figure 
reveals consistent upward trends in ROUGE-L, ROUGE-1, and ROUGE-2 for most models within the early iterations. \texttt{Original LLM set} exhibits the most stable and effective self-improvement behavior, quickly reaching and maintaining high scores. Other models such as \texttt{Mistral-7B-Ins\-truct-v0.3} and \texttt{GPT-3.5-turbo} also improve over iterations, though with more fluctuations. \texttt{Llama-3.2-3B-Instruct} and \texttt{Gemma-2-2b-it} show limited progress, suggesting that lower-capacity models may struggle to utilize refined prompts effectively. This pattern validates the effectiveness of our prompt refinement loop: after each round, \TA leverages feedback (including ROUGE scores and summary outputs) to update the prompt for \SA, leading to consistent performance gains. The early stopping mechanism also helps terminate training for stagnant samples, reducing redundant computation.

The cross-model evaluation demonstrates that \ME achieves reasonable performance across a wide spectrum of LLMs, including lightweight open-source models and high-capacity proprietary systems. This implies 
that the framework is not bound to a specific engine, but rather can flexibly adapt to available resources and deployment contexts.

\vspace{.2cm}
\noindent\fbox{\begin{minipage}{0.98\columnwidth}
		\paragraph{\textbf{Answer to RQ$_3$:}} Using \texttt{Original LLM set} as the agent engine 
		helps \ME obtain a superior recommendation performance. The self-improvement process is effective across the datasets, with most models showing iterative gains in ROUGE scores. Higher-capacity LLMs benefit more from the prompt refinement mechanism, but even smaller models improve initially, validating the robustness of our training loop.
		
\end{minipage}}

%

%
%

	\section{Discussion}	
	\label{sec:Discussion}
	
This section provides discussion related to the implications and possible extensions of our work, as well as the threats to validity of the findings.







	

\subsection{Implications}
Our experimental results provide several implications for both research and practice as follows. 

\vspace{.2cm}
\noindent
\textbf{Usage.} The findings demonstrate that \ME, as a multi-agent framework, consistently outperforms various single-agent LLMs in summarization tasks. This suggests that the collaborative Teacher--Student refinement loop and the dynamic stopping strategy are effective mechanisms for generating concise and faithful summaries. 
By assigning specialized roles to different agents and enabling iterative refinement, \ME is able to generate summaries that are both more accurate and more faithful to the original content. This indicates that the benefit of multi-agent systems lies not only in diversity of models but also in the structured cooperation among them. In practice, such a system could be integrated into SE workflows to automatically produce short descriptions of lengthy documentation, thereby reducing cognitive load for developers and end users.

\vspace{.2cm}
\noindent
\textbf{Use cases.} While our evaluation focuses on Markdown 
files and application descriptions, the approach could potentially be extended to other domains. For example, the framework could support software documentation summarization, where lengthy requirements or technical reports are converted into concise and accessible descriptions. It will also enhance developer support platforms, such as automatically generating short ``About'' sections for repositories or applications, improving accessibility for first-time users. Furthermore, it might benefit community-driven Q\&A platforms by condensing verbose answers into shorter and more searchable forms.

\vspace{.2cm}
\noindent
\textbf{Broader implications.} Beyond the specific task of summarizing \RM and application descriptions, our findings suggest that LLM-based multi-agent systems could serve as a general approach to improve the reliability, scalability, and controllability of AI-assisted summarization. The principles of dynamic prompt refinement and agent specialization can also inspire solutions in adjacent tasks such as code recommendation, bug report summarization, and automated knowledge management.

\subsection{Concerns and Limitations}

While the empirical evaluation demonstrates the effectiveness of \ME, there are various concerns and limitations as follows. 

\begin{itemize}
	
	\item \textbf{Diversity of datasets and application contexts.} The evaluation primarily focused on \GH \RM files and Google Play application descriptions. Although these are representative sources of software documentation, they may not capture the full diversity of real-world documents such as technical reports, requirement specifications, or bug reports. Consequently, the generalizability of \ME to heterogeneous domains still requires further investigation. 
	
	\item  \textbf{Task-oriented and user-centered validation.} Our 
	evaluation relied on benchmark datasets and quantitative metrics (\eg ROUGE, cosine similarity). While these measures have been widely used, they may not fully reflect the utility of summaries in practical scenarios where developers, 
	or end-users consume documentation. Conducting user studies, for instance through crossover experiments comparing human- vs. agent-generated summaries, would provide a deeper understanding of \ME's real-world impact. 
	
	\item  \textbf{Dependence on training data and potential biases.} Although our dataset curation aimed to ensure quality and diversity, the reliance on app descriptions from the Google Play ecosystem may introduce inherent biases, such as domain-specific language or stylistic conventions. Moreover, overlaps with LLM pretraining corpora could unintentionally inflate performance, raising concerns about overfitting. To address this, future evaluations should incorporate datasets from independent sources, including industrial documentation and student-generated artifacts. 
	
	\item  \textbf{Computational cost and scalability.} 
	Despite the efficiency gains achieved by the dynamic iteration strategy, the orchestration of multiple LLM-based agents inevitably introduces computational overhead. Scaling \ME to larger datasets or integrating it into production pipelines may require further optimization or hybrid strategies to balance accuracy and resource consumption. 
	
\end{itemize}

\subsection{Future Research Directions}

Future work on \ME can be extended along several promising directions:

\begin{itemize}
    \item \textbf{Expanding application domains.} While the current evaluation has focused on Markdown files and application descriptions, future work could apply \ME to additional domains such as code summarization, technical documentation, engineering reports, or biomedical texts. Such extensions would allow the framework to demonstrate its adaptability to highly diverse and domain-specific documents.
    
    \item \textbf{Reinforcement learning from feedback.} Another direction is to incorporate reinforcement learning with human feedback (RLHF) or agent feedback (RLAIF) to optimize the prompt refinement and summarization process. This could provide stronger alignment with user expectations and continuously improve the quality of generated summaries.
    
    \item \textbf{Retrieval-augmented generation.} Integrating external retrieval mechanisms into the multi-agent loop would allow agents to access contextual information beyond the input document. This hybrid approach could reduce the risk of missing critical background knowledge and enhance factual accuracy.
    
    \item \textbf{Cost-aware optimization.} Future work should also explore efficient allocation of computational resources by combining lightweight open-source LLMs with larger proprietary ones in a hybrid manner. Such resource-aware scheduling could reduce cost while sustaining high-quality results, making the system more practical in production environments.
    
    \item \textbf{Explainability and transparency.} Finally, improving interpretability by designing logging, tracing, and decision-explanation mechanisms for each agent could strengthen trust in industrial deployments. Providing insights into why a specific summary was produced would enhance both reliability and accountability in real-world software engineering workflows.
\end{itemize}

\subsection{Threats to Validity}

\begin{itemize}

	\item \textbf{Internal Validity} is related to the factors within the experimental design that may bias or distort the results. In our approach, the Teacher--Student refinement loop and dynamic stopping strategy are sensitive to hyperparameters such as iteration thresholds and ROUGE cutoffs. Suboptimal settings might have influenced convergence speed and output quality. 
	Another internal factor is the selection of baseline models: stronger or weaker baselines may lead to different perceived relative improvements. Moreover, single-agent baselines are highly sensitive to model setup (e.g., prompt design, hyperparameter configuration), which may lead to variability in their performance and affect the fairness of comparison. Future experiments should incorporate cross-validation with randomized seeds, multiple runs per configuration, and more systematic baselines to minimize such risks.
	
	\item \textbf{External Validity} concerns the generalizability of our findings to other domains, document types, and usage contexts. While \GH \RM and Google Play descriptions are representative forms of software documentation, they do not capture the broader diversity of real-world documents such as industrial requirement specifications, technical reports, or bug-tracking entries. The language style, structural heterogeneity, and domain-specific jargon in these contexts may pose additional challenges not addressed by our current evaluation. Similarly, the experiments were limited to English-language datasets; generalizing to multilingual documentation or low-resource languages remains unexplored. 

	\item \textbf{Construct Validity} is whether the employed measures and datasets adequately capture the concept of summarization quality. In our evaluation, we relied primarily on ROUGE and cosine similarity metrics, which are well-established but may fail to reflect semantic fidelity, factual consistency, or user-perceived usefulness of the generated summaries. For instance, a summary with high n-gram overlap may still be misleading in meaning. Additionally, we adopted the short ``About'' descriptions from Google Play as ground truth references; however, such descriptions are often written with a marketing or promotional purpose, rather than serving as strictly informative summaries. This introduces the risk that models learn to mimic stylistic tendencies rather than optimizing for clarity or completeness. While we attempted to mitigate this by curating high-quality subsets and applying filtering thresholds, human-centered evaluations (\eg user studies with developers or app users) would provide a more comprehensive assessment of summary utility.


\end{itemize}

	\section{Related Work}
	\label{sec:RelatedWork}



Recently, there has been a surge in the applications of LLM-based MAS in Software Engineering. This section reviews some of the most notable applications, 
as well as MAS for summarization and prompt optimization in agent collaborations.

\subsection{LLM-based Multi-Agent Systems in Software Engineering}
A number of representative applications highlight the potential of LLM-based multi-agent systems (LaMAS) across the SE lifecycle. In program synthesis, systems like \textsc{CodePori}~\citep{rasheed2024codepori} and \textsc{AgentMesh}~\citep{khanzadeh2025agentmesh} demonstrate how assigning specialized roles to multiple agents—such as manager, developers, reviewers, and verification agents--enables the generation of functional, production-level software directly from natural language requirements. \textsc{CodePori} shows strong performance on the \textit{HumanEval} benchmark and further scales to end-to-end applications spanning thousands of lines of code, while \textsc{AgentMesh} showcases artifact-centric communication and iterative plan–code–review loops for reliable development. Moving beyond coding, \textsc{MAGIS}~\citep{tao2024magis} addresses repository-level issue resolution on GitHub by coordinating multiple agents to locate files, handle long contexts, and implement complex fixes, significantly outperforming single-agent LLMs on the \textit{SWE-bench} benchmark.

In industrial contexts, \textsc{GoNoGo}~\citep{Khoee2024GoNoGo} introduces a planner-actor setup to support automotive software release decisions, already deployed at Volvo to improve accuracy and reduce manual overhead in risk-sensitive processes. Similarly, \textsc{CogniSim}~\citep{cinkusz2025agile} integrates cognitive agents into Agile and DevOps workflows, showing improvements in backlog refinement, testing, and deployment, though it also raises questions about transparency and human–agent collaboration. At higher abstraction levels, frameworks such as \textsc{MAAD}~\citep{li2025maad} and multi-agent design/refactoring models automate architecture generation, design trade-offs, and quality balancing by assigning domain-specialized agents (e.g., performance, security, UI/UX) and employing structured evaluation protocols.

LaMAS is emerging as a promising way to streamline software engineering--from requirements to deployment--while also highlighting challenges such as scalability, reliability, security, and the role of human oversight in collaborative, intelligent development.

\subsection{Summarization} 

\ME exemplifies the growing shift toward multi-agent systems (MAS) to enhance the quality, factual consistency, and controllability of LLM-based summarization. By assigning specialized roles, \ie \EA, \SA, and \TA, \ME decomposes the summarization task into modular subtasks with tailored prompts. This role-based specialization is complemented by adaptive strategies, including early halting of low-potential samples and real-time prompt adjustments, enabling efficient and high-quality generation.
This design builds upon prior MAS approaches like MAMM-RE\-FINE~\citep{pan2024mammrefine}, which introduced a \textsc{Detect}-\textsc{Critique}-\textsc{Refine} pipeli\-ne where collaborative agents iteratively improve summaries. Notably, its Multi-Agent Single-Model (MASM) configuration with reranking significantly outperforms single-agent baselines. Metagente extends these ideas by integrating adaptive prompt control and dynamic role coordination throughout the generation process.

Other MAS-based summarization systems further demonstrate the versatility of collaborative agents. Chain-of-Agents~\citep{zhang2024chainofagents} processes long-form input via chained agents under manager oversight, while D\&R~\citep{zhou2025debate} distills stronger models through structured debates and preference optimization. MADRA~\citep{wang2023apollo} enhances agent reasoning with retrieval-augmented evidence to reduce hallucination, and SR-DCR~\citep{zhou2025selfreflective} leverages asymmetric debates and token-level confidence for arbitration. Even in adjacent domains, such as legal argumentation, the Reflective Multi-Agent framework~\citep{zhang2025reflective} highlights the benefits of iterative role-driven collaboration for improving factual and ethical integrity.

Beyond summarization, several MAS frameworks illustrate the power of structured agent collaboration across broader reasoning and task-solving settings. OPRO~\citep{yang2024opro} employs different LLMs as optimizers, using natural language prompts to iteratively generate and refine solutions. APE~\citep{zhou2023ape} adopts a similar philosophy by generating candidate instructions and refining them via semantic similarity and evaluation metrics. Camel~\citep{li2024camel} introduces a role-playing framework where agents are guided by inception prompting, enabling instruction-following cooperation and generation of multi-agent conversational data. \texttt{MetaGPT}~\citep{hong2024metagpt} embeds human workflows into modular agent behaviors to address hallucination and coordination challenges. Complementary to these systems, \texttt{AutoGen}~\citep{wu2024autogen} and \texttt{LangChain}\footnote{\url{https://docs.langchain.com}} facilitate the development of LLM-based applications via conversation-driven agents and modular programming abstractions. \texttt{AutoGen} promotes human-in-the-loop interactions through multi-turn dialog and agent orchestration, while \texttt{LangChain} streamlines integration with external tools and data sources.

Altogether, these studies contextualize and reinforce \ME's contributions: dynamic agent specialization, modular workflows, and adaptive control as key mechanisms for building faithful and effective multi-agent summarization systems.

\subsection{Prompt Optimization in Agent Collaboration}

\ME emphasizes prompt optimization as a core mechanism for effective multi-agent collaborations. Its architecture incorporates dynamic, role-specific prompt refinement—most notably via \TA, which continuously updates summarization prompts based on ROUGE-L feedback. This closed-loop adaptation enables agents to specialize and respond to diverse inputs with increased coherence, factuality, and control.
This design is in line with broader efforts to optimize prompts across agent-based systems. MASS~\citep{zhou2025mass} introduces a framework that jointly searches for both communication topologies and agent prompts, demonstrating that interleaving local and global prompt tuning leads to better system-wide coordination. Similarly, NEXUSSUM~\citep{kim2025nexussum} applies modular prompt designs across hierarchical agent roles--Preprocessor, Summarizer, and Compressor—using Chain-of-Thought reasoning and few-shot examples to guide generation.

Further advances in Automated Prompt Optimization (APO) support the principles underpinning \ME. MARS~\citep{zhang2025mars} proposes a multi-agent dialogue framework, where agents like Planner and Teacher-Critic-Student collaboratively evolve prompts. Shen et al.~\citep{shen2025optimizing} explored feedback-driven refinement, showing that group-level, online optimization consistently improves collaboration quality. Meanwhile, SEE~\citep{cui2025see} and EXPO~\citep{kong2025meta} treated prompt tuning as a high-dimensional or dynamic problem, applying metaheuristics and bandit learning to refine both instructions and exemplars.

Collectively, these studies reinforce the design choices in \ME: prompt modularity, agent-level specialization, and adaptive optimization loops are essential to building robust, controllable LLM-based multi-agent systems.
	\section{Conclusion and Future Work}
	\label{sec:Conclusions}

This paper presented \ME--a novel approach to summarization of software documents leveraging LLM-based multi agents. Using datasets collected for mobile apps' documents, we conducted an empirical evaluation to study the performance of \ME. Moreover, we also compared dynamic vs. non-dynamic strategies, evaluating generalization across test subsets. The experimental results showed that \ME is able to generate highly relevant summaries, outperforming single LLMs. Our work showcases the potential of LLM-based MAS in the generation of software documents. 

Future work will focus on the deployment of the proposed architecture to a wider range of application domains, beyond the scenarios investigated in this study. In particular, we plan to explore tasks such as code recommendation, automated code comment generation, and other software engineering activities where collaborative LLM-based agents can provide added value. A key challenge in these extensions will be to refine and optimize the architecture of the LLM-based MAS, with the dual objective of improving the overall effectiveness of the system and reducing its computational cost, thus making it more suitable for real-world integration in developer workflows.

In addition, an important research direction concerns the definition of benchmarking framework tailored to LLM-based MAS in Software Engineering. Such a framework should not only enable rigorous comparisons across different agent architectures and coordination strategies, but also include metrics that capture aspects such as accuracy, efficiency, robustness, scalability, and fairness. Establishing these benchmarks will be instrumental for the community to assess progress in this emerging area and to guide the development of more reliable, efficient, and trustworthy multi-agent solutions.
%





%
%

	\section*{Declarations}
	
	 \textbf{Conflict of interest.} All the authors declare that they have no conflict of interest. Furthermore, they have no known competing financial interests or personal relationships that could have appeared to influence the work reported in this paper.
	 
	 \vspace{.2cm}
	 \noindent
	 \textbf{Ethical approval.} This article does not contain any studies with human  participants or animals performed by any of the authors.
	
	
%
%
%
%
%
%
%
%
%


		
%

	\begin{acknowledgements}
		This paper has been partially supported by the MOSAICO project (Management, Orchestration and Supervision of AI-agent COmmunities for reliable AI in software engineering) that has received funding from the European Union under the Horizon Research and Innovation Action (Grant Agreement No. 101189664). 
		The work has been also partially supported by the European Union--NextGenerationEU through the Italian Ministry of University and Research, Projects PRIN 2022 PNRR \emph{``FRINGE: context-aware FaiRness engineerING in complex software systEms''} grant n. P2022553SL. We acknowledge the Italian ``PRIN 2022'' project TRex-SE: \emph{``Trustworthy Recommenders for Software Engineers,''} grant n. 2022LKJWHC.
	\end{acknowledgements}

	%
	%

	\bibliographystyle{abbrvnat}

\bibliography{main}

@article{sun2024source,
	title={Source code summarization in the era of large language models},
	author={Sun, Weisong and Miao, Yun and Li, Yuekang and Zhang, Hongyu and Fang, Chunrong and Liu, Yi and Deng, Gelei and Liu, Yang and Chen, Zhenyu},
	journal={arXiv preprint arXiv:2407.07959},
	year={2024}
}

@inproceedings{haldar2024analyzing,
	title={Analyzing the Performance of Large Language Models on Code Summarization},
	author={Haldar, Rajarshi and Hockenmaier, Julia},
	booktitle={Proceedings of the 2024 Joint International Conference on Computational Linguistics, Language Resources and Evaluation (LREC-COLING 2024)},
	pages={995--1008},
	year={2024}
}

@inproceedings{lee2024github,
	title={The github recent bugs dataset for evaluating llm-based debugging applications},
	author={Lee, Jae Yong and Kang, Sungmin and Yoon, Juyeon and Yoo, Shin},
	booktitle={2024 IEEE Conference on Software Testing, Verification and Validation (ICST)},
	pages={442--444},
	year={2024},
	organization={IEEE}
}

@inproceedings{tian2024debugbench,
	title={DebugBench: Evaluating Debugging Capability of Large Language Models},
	author={Tian, Runchu and Ye, Yining and Qin, Yujia and Cong, Xin and Lin, Yankai and Pan, Yinxu and Wu, Yesai and Haotian, Hui and Weichuan, Liu and Liu, Zhiyuan and others},
	booktitle={Findings of the Association for Computational Linguistics ACL 2024},
	pages={4173--4198},
	year={2024}
}

@inproceedings{arawjo2024chainforge,
	title={Chainforge: A visual toolkit for prompt engineering and llm hypothesis testing},
	author={Arawjo, Ian and Swoopes, Chelse and Vaithilingam, Priyan and Wattenberg, Martin and Glassman, Elena L},
	booktitle={Proceedings of the 2024 CHI Conference on Human Factors in Computing Systems},
	pages={1--18},
	year={2024}
}

@article{li2025enhancing,
	title={Enhancing Differential Testing With LLMs For Testing Deep Learning Libraries},
	author={Li, Meiziniu and Li, Dongze and Liu, Jianmeng and Cao, Jialun and Tian, Yongqiang and Cheung, Shing-Chi},
	journal={ACM Transactions on Software Engineering and Methodology},
	year={2025},
	publisher={ACM New York, NY}
}

@inproceedings{gu2023llm,
	title={Llm-based code generation method for golang compiler testing},
	author={Gu, Qiuhan},
	booktitle={Proceedings of the 31st ACM Joint European Software Engineering Conference and Symposium on the Foundations of Software Engineering},
	pages={2201--2203},
	year={2023}
}

@article{fakhoury2024llm,
	title={Llm-based test-driven interactive code generation: User study and empirical evaluation},
	author={Fakhoury, Sarah and Naik, Aaditya and Sakkas, Georgios and Chakraborty, Saikat and Lahiri, Shuvendu K},
	journal={IEEE Transactions on Software Engineering},
	year={2024},
	publisher={IEEE}
}

@article{huang2024bias,
	title={Bias testing and mitigation in llm-based code generation},
	author={Huang, Dong and Zhang, Jie M and Bu, Qingwen and Xie, Xiaofei and Chen, Junjie and Cui, Heming},
	journal={ACM Transactions on Software Engineering and Methodology},
	year={2024},
	publisher={ACM New York, NY}
}

@article{LIU2017126,
	title = {Mining domain knowledge from app descriptions},
	journal = {Journal of Systems and Software},
	volume = {133},
	pages = {126-144},
	year = {2017},
	issn = {0164-1212},
	doi = {https://doi.org/10.1016/j.jss.2017.08.024},
	url = {https://www.sciencedirect.com/science/article/pii/S0164121217301784},
	author = {Yuzhou Liu and Lei Liu and Huaxiao Liu and Xiaoyu Wang and Hongji Yang}
}

@article{LIU2022106924,
	title = {How ReadMe files are structured in open source Java projects},
	journal = {Information and Software Technology},
	volume = {148},
	pages = {106924},
	year = {2022},
	issn = {0950-5849},
	doi = {https://doi.org/10.1016/j.infsof.2022.106924},
	url = {https://www.sciencedirect.com/science/article/pii/S0950584922000775},
	author = {Yuyang Liu and Ehsan Noei and Kelly Lyons}
}

@article{BORGES2018112,
	title = {{What’s in a GitHub Star? Understanding Repository Starring Practices in a Social Coding Platform}},
	journal = {Journal of Systems and Software},
	volume = {146},
	pages = {112-129},
	year = {2018},
	issn = {0164-1212},
	doi = {https://doi.org/10.1016/j.jss.2018.09.016},
	url = {https://www.sciencedirect.com/science/article/pii/S0164121218301961},
	author = {Hudson Borges and Marco {Tulio Valente}}
}

@inproceedings{10.1145/3593434.3593448,
	author = {Doan, Thu T. H. and Nguyen, Phuong T. and Di Rocco, Juri and Di Ruscio, Davide},
	title ={{Too long; didn’t read: Automatic summarization of GitHub README.MD with Transformers}},
	year = {2023},
	isbn = {9798400700446},
	publisher = {Association for Computing Machinery},
	address = {New York, NY, USA},
	url = {https://doi.org/10.1145/3593434.3593448},
	doi = {10.1145/3593434.3593448},
	booktitle = {Proceedings of the 27th International Conference on Evaluation and Assessment in Software Engineering},
	pages = {267–272},
	numpages = {6},
	location = {Oulu, Finland},
	series = {EASE '23}
}

@inproceedings{10.1145/3691620.3695291,
	author = {Wang, Luqiao and Zhou, Yangtao and Zhuang, Huiying and Li, Qingshan and Cui, Di and Zhao, Yutong and Wang, Lu},
	title = {Unity Is Strength: Collaborative LLM-Based Agents for Code Reviewer Recommendation},
	year = {2024},
	isbn = {9798400712487},
	publisher = {ACM},
	address = {New York, NY, USA},
	url = {https://doi.org/10.1145/3691620.3695291},
	doi = {10.1145/3691620.3695291},
	booktitle = {Proceedings of the 39th IEEE/ACM ASE},
	pages = {2235–2239},
	numpages = {5},
	series = {ASE '24}
}

@inproceedings{white2023promptpatterncatalogenhance,
	author = {White, Jules and Fu, Quchen and Hays, Sam and Sandborn, Michael and Olea, Carlos and Gilbert, Henry and Elnashar, Ashraf and Spencer-Smith, Jesse and Schmidt, Douglas C.},
	title = {A Prompt Pattern Catalog to Enhance Prompt Engineering with ChatGPT},
	year = {2023},
	isbn = {9781941652190},
	publisher = {The Hillside Group},
	address = {USA},
	booktitle = {Proceedings of the 30th Conference on Pattern Languages of Programs},
	articleno = {5},
	numpages = {31},
	location = {Monticello, IL, USA},
	series = {PLoP '23}
}

@article{DBLP:journals/corr/abs-2407-01489,
	author       = {Chunqiu Steven Xia and
	Yinlin Deng and
	Soren Dunn and
	Lingming Zhang},
	title        = {Agentless: Demystifying LLM-based Software Engineering Agents},
	journal      = {CoRR},
	year         = {2024},
	url          = {https://doi.org/10.48550/arXiv.2407.01489},
	doi          = {10.48550/ARXIV.2407.01489},
	eprinttype    = {arXiv},
	eprint       = {2407.01489},
	bibsource    = {dblp computer science bibliography, https://dblp.org}
}

@article{DBLP:journals/fcsc/WangMFZYZCTCLZWW24,
	author       = {Lei Wang and
	Chen Ma and
	Xueyang Feng and
	Zeyu Zhang and
	Hao Yang and
	Jingsen Zhang and
	Zhiyuan Chen and
	Jiakai Tang and
	Xu Chen and
	Yankai Lin and
	Wayne Xin Zhao and
	Zhewei Wei and
	Jirong Wen},
	title        = {A survey on large language model based autonomous agents},
	journal      = {Frontiers Comput. Sci.},
	volume       = {18},
	number       = {6},
	pages        = {186345},
	year         = {2024},
	url          = {https://doi.org/10.1007/s11704-024-40231-1},
	doi          = {10.1007/S11704-024-40231-1},
	bibsource    = {dblp computer science bibliography, https://dblp.org}
}

@article{He_Treude_Lo_2024,
	author = {He, Junda and Treude, Christoph and Lo, David},
	title = {LLM-Based Multi-Agent Systems for Software Engineering: Literature Review, Vision and the Road Ahead},
	year = {2025},
	publisher = {Association for Computing Machinery},
	address = {New York, NY, USA},
	issn = {1049-331X},
	url = {https://doi.org/10.1145/3712003},
	doi = {10.1145/3712003},
	note = {Just Accepted},
	journal = {ACM Trans. Softw. Eng. Methodol.},
}

@article{NGUYEN2024112059,
	title = {{GPTSniffer: A CodeBERT-based classifier to detect source code written by ChatGPT}},
	journal = {Journal of Systems and Software},
	volume = {214},
	pages = {112059},
	year = {2024},
	issn = {0164-1212},
	doi = {https://doi.org/10.1016/j.jss.2024.112059},
	url = {https://www.sciencedirect.com/science/article/pii/S0164121224001043},
	author = {Phuong T. Nguyen and Juri {Di Rocco} and Claudio {Di Sipio} and Riccardo Rubei and Davide {Di Ruscio} and Massimiliano {Di Penta}}
}

@article{DBLP:journals/software/Ozkaya23b,
	author       = {Ipek Ozkaya},
	title        = {Application of Large Language Models to Software Engineering Tasks:
	Opportunities, Risks, and Implications},
	journal      = {{IEEE} Softw.},
	volume       = {40},
	number       = {3},
	pages        = {4--8},
	year         = {2023},
	url          = {https://doi.org/10.1109/MS.2023.3248401},
	doi          = {10.1109/MS.2023.3248401},
	bibsource    = {dblp computer science bibliography, https://dblp.org}
}

@inproceedings{10.1145/3696630.3728511,
	author = {Nguyen, Duc S. H. and Truong, Bach G. and Nguyen, Phuong T. and Di Rocco, Juri and Di Ruscio, Davide},
	title = {{Teamwork makes the dream work: LLMs-Based Agents for GitHub README.MD Summarization}},
	year = {2025},
	isbn = {9798400712760},
	publisher = {Association for Computing Machinery},
	address = {New York, NY, USA},
	url = {https://doi.org/10.1145/3696630.3728511},
	doi = {10.1145/3696630.3728511},
	booktitle = {Proceedings of the 33rd ACM International Conference on the Foundations of Software Engineering},
	pages = {621–625},
	numpages = {5},
	location = {Clarion Hotel Trondheim, Trondheim, Norway},
	series = {FSE Companion '25}
}

@inproceedings{rouge2004,
	title = "{ROUGE}: A Package for Automatic Evaluation of Summaries",
	author = "Lin, Chin-Yew",
	booktitle = "Text Summarization Branches Out",
	month = jul,
	year = "2004",
	address = "Barcelona, Spain",
	publisher = "Association for Computational Linguistics",
	url = "https://aclanthology.org/W04-1013",
	pages = "74--81",
}

@inproceedings{Chen2020,
	author = {Chen, Songqiang and Xie, Xiaoyuan and Yin, Bangguo and Ji, Yuanxiang and Chen, Lin and Xu, Baowen},
	title = {Stay Professional and Efficient: Automatically Generate Titles for Your Bug Reports},
	year = {2021},
	isbn = {9781450367684},
	publisher = {Association for Computing Machinery},
	address = {New York, NY, USA},
	url = {https://doi.org/10.1145/3324884.3416538},
	doi = {10.1145/3324884.3416538},
	booktitle = {Proceedings of the 35th IEEE/ACM International Conference on Automated Software Engineering},
	pages = {385–397},
	numpages = {13},
	location = {Virtual Event, Australia},
	series = {ASE '20}
}

@inproceedings{itiger2022,
	author = {Zhang, Ting and Irsan, Ivana Clairine and Thung, Ferdian and Han, DongGyun and Lo, David and Jiang, Lingxiao},
	title = {iTiger: An Automatic Issue Title Generation Tool},
	year = {2022},
	isbn = {9781450394130},
	publisher = {Association for Computing Machinery},
	address = {New York, NY, USA},
	url = {https://doi.org/10.1145/3540250.3558934},
	doi = {10.1145/3540250.3558934},
	pages = {1637–1641},
	numpages = {5},
	location = {Singapore, Singapore},
	series = {ESEC/FSE 2022}
}

@inproceedings{Allix:2016:ACM:2901739.2903508,
	author = {Allix, Kevin and Bissyand{\'e}, Tegawend{\'e} F. and Klein, Jacques and Le Traon, Yves},
	title = {AndroZoo: Collecting Millions of Android Apps for the Research Community},
	booktitle = {Proceedings of the 13th International Conference on Mining Software Repositories},
	series = {MSR '16},
	year = {2016},
	isbn = {978-1-4503-4186-8},
	location = {Austin, Texas},
	pages = {468--471},
	numpages = {4},
	url = {http://doi.acm.org/10.1145/2901739.2903508},
	doi = {10.1145/2901739.2903508},
	acmid = {2903508},
	publisher = {ACM},
	address = {New York, NY, USA},
}

@article{pan2024mammrefine,
  title     = {MAMM-REFINE: Faithful Multi-Agent Summarization with Automatic Mistake Mining and Self-Refinement},
  author    = {Liangming Pan and Yuxiang Wu and Yiqun Liu and Maosong Sun and Yeyun Gong and Zhiyi Yang and Daxin Jiang},
  journal   = {arXiv preprint arXiv:2503.15272},
  year      = {2024},
  eprint    = {2503.15272},
  archivePrefix = {arXiv},
  primaryClass  = {cs.CL},
  url       = {https://arxiv.org/abs/2503.15272}
}

@inproceedings{zhang2025reflective,
  title     = {Mitigating Manipulation and Enhancing Persuasion: A Reflective Multi-Agent Approach for Legal Argument Generation},
  author    = {Li Zhang and Kevin D. Ashley},
  booktitle = {Proceedings of the Workshop on Legally Compliant Intelligent Chatbots at ICAIL 2025},
  year      = {2025},
  url       = {https://arxiv.org/abs/2506.02992},
  archivePrefix = {arXiv},
  eprint    = {2506.02992},
  primaryClass = {cs.AI},
  address   = {Chicago, IL, USA}
}

@article{zhou2025selfreflective,
  title     = {When to Trust Context: Self-Reflective Debates for Context Reliability},
  author    = {Zeqi Zhou and Fang Wu and Shayan Talaei and Haokai Zhao and Cheng Meixin and Tinson Xu and Amin Saberi and Yejin Choi},
  journal   = {arXiv preprint arXiv:2506.06020},
  year      = {2025},
  eprint    = {2506.06020},
  archivePrefix = {arXiv},
  primaryClass  = {cs.CL},
  url       = {https://arxiv.org/abs/2506.06020}
}

@article{zhang2024chainofagents,
  title     = {Chain of Agents: Large Language Models Collaborating on Long-Context Tasks},
  author    = {Yusen Zhang and Ruoxi Sun and Yanfei Chen and Tomas Pfister and Rui Zhang and Sercan \"O. Arik},
  journal   = {arXiv preprint arXiv:2406.02818},
  year      = {2024},
  eprint    = {2406.02818},
  archivePrefix = {arXiv},
  primaryClass  = {cs.CL},
  url       = {https://arxiv.org/abs/2406.02818},
  note      = {Preprint. Under review. Work done in part at Google Cloud AI Research.}
}

@article{zhou2025debate,
  title={Debate, Reflect, and Distill: Multi-Agent Feedback with Tree-Structured Preference Optimization for Efficient Language Model Enhancement},
  author={Xiaofeng Zhou and Heyan Huang and Lizi Liao},
  journal={arXiv preprint arXiv:2506.03541},
  year={2025},
  url={https://arxiv.org/abs/2506.03541}
}

@article{wang2023apollo,
  title     = {Apollo’s Oracle: Retrieval-Augmented Reasoning in Multi-Agent Debates},
  author    = {Haotian Wang and Xiyuan Du and Weijiang Yu and Qianglong Chen and Kun Zhu and Zheng Chu and Lian Yan and Yi Guan},
  journal   = {arXiv preprint arXiv:2312.04854},
  year      = {2023},
  eprint    = {2312.04854},
  archivePrefix = {arXiv},
  primaryClass  = {cs.CL},
  url       = {https://arxiv.org/abs/2312.04854}
}

@article{zhou2025mass,
  title     = {Multi-Agent Design: Optimizing Agents with Better Prompts and Topologies},
  author    = {Zhou, Han and Wan, Xingchen and Sun, Ruoxi and Palangi, Hamid and Iqbal, Shariq and Vuli{\'c}, Ivan and Korhonen, Anna and Ar{\i}k, Sercan {\"O}},
  journal   = {arXiv preprint arXiv:2502.02533},
  year      = {2025},
  month     = {February},
  eprint    = {2502.02533},
  archivePrefix = {arXiv},
  primaryClass  = {cs.LG},
  institution = {Google Research and University of Cambridge},
  url       = {https://arxiv.org/abs/2502.02533},
  note      = {Preprint submitted to arXiv}
}

@misc{kim2025nexussum,
  title={NEXUSSUM: Hierarchical LLM Agents for Long-Form Narrative Summarization},
  author={Hyuntak Kim and Byung-Hak Kim},
  year={2025},
  eprint={2505.24575},
  archivePrefix={arXiv},
  primaryClass={cs.CL},
  url={https://arxiv.org/abs/2505.24575}
}

@misc{zhang2025mars,
  title={MARS: A Multi-Agent Framework Incorporating Socratic Guidance for Automated Prompt Optimization},
  author={Jian Zhang and Zhangqi Wang and Haiping Zhu and Jun Liu and Qika Lin and Erik Cambria},
  year={2025},
  eprint={2503.16874},
  archivePrefix={arXiv},
  primaryClass={cs.CL},
  url={https://arxiv.org/abs/2503.16874}
}

@inproceedings{shen2025optimizing,
  title={Optimizing LLM-Based Multi-Agent System with Textual Feedback: A Case Study on Software Development},
  author={Shen, Ming and Shu, Raphael and Pratik, Anurag and Gung, James and Ge, Yubin and Sunkara, Monica and Zhang, Yi},
  booktitle={AI Agents: Capabilities and Safety Workshop @ Conference on Language Modeling (COLM)},
  year={2025},
  url={https://arxiv.org/abs/2505.16086}
}

@article{cui2025see,
  title={SEE: Strategic Exploration and Exploitation for Cohesive In-Context Prompt Optimization},
  author={Cui, Wendi and Li, Zhuohang and Sun, Hao and Lopez, Damien and Das, Kamalika and Malin, Bradley and Kumar, Sricharan and Zhang, Jiaxin},
  journal={arXiv preprint arXiv:2402.11347},
  year={2025},
  url={https://arxiv.org/abs/2402.11347}
}

@article{kong2025meta,
  title={Meta-Prompt Optimization for LLM-Based Sequential Decision Making},
  author={Kong, Mingze and Wang, Zhiyong and Shu, Yao and Dai, Zhongxiang},
  journal={arXiv preprint arXiv:2502.00728},
  year={2025},
  url={https://arxiv.org/abs/2502.00728}
}

@inproceedings{yang2024opro,
  title     = {Large Language Models as Optimizers},
  author    = {Chengrun Yang and Xuezhi Wang and Yifeng Lu and Hanxiao Liu and Quoc V. Le and Denny Zhou and Xinyun Chen},
  booktitle = {Proceedings of the Twelfth International Conference on Learning Representations (ICLR)},
  year      = {2024},
  address   = {Vienna, Austria},
  url       = {https://openreview.net/forum?id=Bb4VGOWELI}
}

@inproceedings{zhou2023ape,
  title     = {Large Language Models are Human-Level Prompt Engineers},
  author    = {Yongchao Zhou and Andrei Ioan Muresanu and Ziwen Han and Keiran Paster and Silviu Pitis and Harris Chan and Jimmy Ba},
  booktitle = {Proceedings of the Eleventh International Conference on Learning Representations (ICLR)},
  year      = {2023},
  address   = {Kigali, Rwanda},
  url       = {https://openreview.net/forum?id=92gvk82DE-}
}

@inproceedings{li2024camel,
  title     = {CAMEL: Communicative Agents for "Mind" Exploration of Large Language Model Society},
  author    = {Guohao Li and Hasan Abed Al Kader Hammoud and Hani Itani and Dmitrii Khizbullin and Bernard Ghanem},
  booktitle = {Proceedings of the 37th International Conference on Neural Information Processing Systems (NeurIPS)},
  year      = {2024},
  address   = {New Orleans, LA, USA},
  publisher = {Curran Associates Inc.},
  note      = {Article 2264, 18 pages}
}

@inproceedings{hong2024metagpt,
  title     = {MetaGPT: Meta Programming for a Multi-Agent Collaborative Framework},
  author    = {Sirui Hong and Mingchen Zhuge and Jonathan Chen and Xiawu Zheng and Yuheng Cheng and Jinlin Wang and Ceyao Zhang and Zili Wang and Steven Ka Shing Yau and Zijuan Lin and Liyang Zhou and Chenyu Ran and Lingfeng Xiao and Chenglin Wu and J{\"u}rgen Schmidhuber},
  booktitle = {Proceedings of the Twelfth International Conference on Learning Representations (ICLR)},
  year      = {2024},
  url       = {https://openreview.net/forum?id=VtmBAGCN7o}
}

@inproceedings{wu2024autogen,
  title     = {AutoGen: Enabling Next-Gen LLM Applications via Multi-Agent Conversations},
  author    = {Qingyun Wu and Gagan Bansal and Jieyu Zhang and Yiran Wu and Beibin Li and Erkang Zhu and Li Jiang and Xiaoyun Zhang and Shaokun Zhang and Jiale Liu and Ahmed Hassan Awadallah and Ryen W. White and Doug Burger and Chi Wang},
  booktitle = {Proceedings of the First Conference on Language Modeling},
  year      = {2024},
  url       = {https://openreview.net/forum?id=BAakY1hNKS}
}

@article{rasheed2024codepori,
  title     = {CodePori: Large-Scale System for Autonomous Software Development Using Multi-Agent Technology},
  author    = {Rasheed, Zeeshan and Sami, Abdul Malik and Kemell, Kai-Kristian and Waseem, Muhammad and Saari, Mika and Syst{\"a}, Kari and Abrahamsson, Pekka},
  journal   = {Information and Software Technology},
  year      = {2024},
  note      = {Preprint, arXiv:2402.01411},
  url       = {https://arxiv.org/abs/2402.01411},
}

@misc{khanzadeh2025agentmesh,
  title={AgentMesh: A Cooperative Multi-Agent Generative AI Framework for Software Development Automation},
  author={Khanzadeh, Sourena},
  year={2025},
  eprint={2507.19902},
  archivePrefix={arXiv},
  primaryClass={cs.SE},
  url={https://arxiv.org/abs/2507.19902v1}
}

@article{Khoee2024GoNoGo,
  title     = {GoNoGo: An Efficient LLM-based Multi-Agent System for Streamlining Automotive Software Release Decision-Making},
  author    = {Arsham Gholamzadeh Khoee and Yinan Yu and Robert Feldt and Andris Freimanis and Patrick Andersson Rhodin and Dhasarathy Parthasarathy},
  journal   = {arXiv preprint arXiv:2408.09785},
  year      = {2024},
  url       = {https://arxiv.org/abs/2408.09785}
}

@inproceedings{cinkusz2025agile,
  author    = {Konrad Cinkusz and Jaros{\l}aw A. Chudziak},
  title     = {Agile Software Management with Cognitive Multi-Agent Systems},
  booktitle = {Proceedings of the 17th International Conference on Agents and Artificial Intelligence (ICAART 2025) - Volume 1},
  pages     = {385--392},
  year      = {2025},
  publisher = {SCITEPRESS},
  doi       = {10.5220/0013153000003890},
  isbn      = {978-989-758-737-5},
  issn      = {2184-433X}
}

@inproceedings{tao2024magis,
  title     = {MAGIS: LLM-Based Multi-Agent Framework for GitHub Issue Resolution},
  author    = {Tao, Wei and Zhou, Yucheng and Wang, Yanlin and Zhang, Wenqiang and Zhang, Hongyu and Cheng, Yu},
  booktitle = {Proceedings of the 38th Conference on Neural Information Processing Systems (NeurIPS)},
  year      = {2024},
  publisher = {Curran Associates, Inc.}
}

@article{li2025maad,
  author    = {Ruiyin Li and Yiran Zhang and Xiyu Zhou and Peng Liang and Weisong Sun and Jifeng Xuan and Zhi Jin and Yang Liu},
  title     = {MAAD: Automate Software Architecture Design through Knowledge-Driven Multi-Agent Collaboration},
  journal   = {ACM Transactions on Software Engineering and Methodology},
  volume    = {0},
  number    = {0},
  pages     = {Article 0},
  year      = {2025},
  publisher = {Association for Computing Machinery},
  doi       = {10.1145/nnnnnnn.nnnnnnn},
  url       = {https://arxiv.org/abs/2507.21382v1}
}

@article{Aly2025,
title = {Cross-Domain Evaluation of Large Language Models for Abstractive Text Summarization: An Empirical Perspective},
journal = {International Journal of Advanced Computer Science and Applications},
doi = {10.14569/IJACSA.2025.0160695},
url = {http://dx.doi.org/10.14569/IJACSA.2025.0160695},
year = {2025},
publisher = {The Science and Information Organization},
volume = {16},
number = {6},
author = {Walid Mohamed Aly and Taysir Hassan A. Soliman and Amr Mohamed AbdelAziz}
}

@inproceedings{takeshita-etal-2025-irsum,
title = "{IRS}um: One Model to Rule Summarization and Retrieval",
author = "Takeshita, Sotaro  and
Ponzetto, Simone Paolo  and
Eckert, Kai",
editor = "Arviv, Ofir  and
Clinciu, Miruna  and
Dhole, Kaustubh  and
Dror, Rotem  and
Gehrmann, Sebastian  and
Habba, Eliya  and
Itzhak, Itay  and
Mille, Simon  and
Perlitz, Yotam  and
Santus, Enrico  and
Sedoc, Jo{\~a}o  and
Shmueli Scheuer, Michal  and
Stanovsky, Gabriel  and
Tafjord, Oyvind",
booktitle = "Proceedings of the Fourth Workshop on Generation, Evaluation and Metrics (GEM{\texttwosuperior})",
month = jul,
year = "2025",
address = "Vienna, Austria and virtual meeting",
publisher = "Association for Computational Linguistics",
url = "https://aclanthology.org/2025.gem-1.23/",
pages = "262--275",
ISBN = "979-8-89176-261-9"
}

@inbook{10.1145/3677389.3702588,
author = {Keya, Farhana and Jaradeh, Mohamad Yaser and Auer, S\'{o}ren},
title = {Leveraging LLMs for Scientific Abstract Summarization: Unearthing the Essence of Research in a Single Sentence},
year = {2025},
isbn = {9798400710933},
publisher = {Association for Computing Machinery},
address = {New York, NY, USA},
url = {https://doi-org.univaq.idm.oclc.org/10.1145/3677389.3702588},
articleno = {9},
numpages = {7}
}

@INPROCEEDINGS{10986332,
	author={Khan, Rimsha and Sharma, Shanu and Upadhyay, Divya},
	booktitle={2025 3rd International Conference on Disruptive Technologies (ICDT)}, 
	title={Extracting Abstractive Summaries Through Generative AI Models}, 
	year={2025},
	volume={},
	number={},
	pages={671-676},
	doi={10.1109/ICDT63985.2025.10986332}
}
\end{document}